\documentclass[%
 reprint,
superscriptaddress,
 amsmath,amssymb,
prx,aps,
showpacs
]{revtex4-2}

\usepackage{graphicx}
\usepackage{dcolumn}
\usepackage{bm}
\usepackage{hyperref}
\usepackage{mathrsfs}
\usepackage{amsthm}
\usepackage{braket}
\usepackage{gensymb}
\usepackage{siunitx}
\usepackage{textcomp}
\usepackage{upgreek}
\usepackage{romannum}
\usepackage{dsfont}
\usepackage{makecell}
\usepackage{multirow}
\usepackage[table]{xcolor}
\usepackage{comment}

\newcommand{\eq}{Eq.~}

\newcommand{\eqs}{Eqs.~}
\newcommand{\fig}{Fig.~}
\newcommand{\figs}{Figs.~}

\newcommand{\sect}{Sec.~}
\def\app{App.~}

\AtBeginDocument{\pagenumbering{arabic}}
\begin{document}

\author{V. Jouanny}
\affiliation{Hybrid Quantum Circuits Laboratory (HQC), Institute of Physics, \'{E}cole Polytechnique F\'{e}d\'{e}rale de Lausanne (EPFL), 1015, Lausanne, Switzerland}
\affiliation{Center for Quantum Science and Engineering,\\ \ Institute of Physics, \'{E}cole Polytechnique F\'{e}d\'{e}rale de Lausanne (EPFL), 1015, Lausanne, Switzerland}
\author{S. Frasca}
\affiliation{Hybrid Quantum Circuits Laboratory (HQC), Institute of Physics, \'{E}cole Polytechnique F\'{e}d\'{e}rale de Lausanne (EPFL), 1015, Lausanne, Switzerland}
\affiliation{Center for Quantum Science and Engineering,\\ \ Institute of Physics, \'{E}cole Polytechnique F\'{e}d\'{e}rale de Lausanne (EPFL), 1015, Lausanne, Switzerland}
\author{V.J. Weibel}%
\affiliation{Hybrid Quantum Circuits Laboratory (HQC), Institute of Physics, \'{E}cole Polytechnique F\'{e}d\'{e}rale de Lausanne (EPFL), 1015, Lausanne, Switzerland}
\affiliation{Center for Quantum Science and Engineering,\\ \ Institute of Physics, \'{E}cole Polytechnique F\'{e}d\'{e}rale de Lausanne (EPFL), 1015, Lausanne, Switzerland}
\author{L. Peyruchat}
\affiliation{Hybrid Quantum Circuits Laboratory (HQC), Institute of Physics, \'{E}cole Polytechnique F\'{e}d\'{e}rale de Lausanne (EPFL), 1015, Lausanne, Switzerland}
\affiliation{Center for Quantum Science and Engineering,\\ \ Institute of Physics, \'{E}cole Polytechnique F\'{e}d\'{e}rale de Lausanne (EPFL), 1015, Lausanne, Switzerland}
\author{M. Scigliuzzo}
\affiliation{Laboratory of Photonics and Quantum Measurements (LPQM), Institute of Physics, \'{E}cole Polytechnique F\'{e}d\'{e}rale de Lausanne (EPFL), 1015, Lausanne, Switzerland}
\affiliation{Center for Quantum Science and Engineering,\\ \ Institute of Physics, \'{E}cole Polytechnique F\'{e}d\'{e}rale de Lausanne (EPFL), 1015, Lausanne, Switzerland}
\author{F. Oppliger}
\affiliation{Hybrid Quantum Circuits Laboratory (HQC), Institute of Physics, \'{E}cole Polytechnique F\'{e}d\'{e}rale de Lausanne (EPFL), 1015, Lausanne, Switzerland}
\affiliation{Center for Quantum Science and Engineering,\\ \ Institute of Physics, \'{E}cole Polytechnique F\'{e}d\'{e}rale de Lausanne (EPFL), 1015, Lausanne, Switzerland}
\author{F. De Palma}
\affiliation{Hybrid Quantum Circuits Laboratory (HQC), Institute of Physics, \'{E}cole Polytechnique F\'{e}d\'{e}rale de Lausanne (EPFL), 1015, Lausanne, Switzerland}
\affiliation{Center for Quantum Science and Engineering,\\ \ Institute of Physics, \'{E}cole Polytechnique F\'{e}d\'{e}rale de Lausanne (EPFL), 1015, Lausanne, Switzerland}
\author{D. Sbroggi\`o}
\affiliation{Hybrid Quantum Circuits Laboratory (HQC), Institute of Physics, \'{E}cole Polytechnique F\'{e}d\'{e}rale de Lausanne (EPFL), 1015, Lausanne, Switzerland}
\affiliation{Center for Quantum Science and Engineering,\\ \ Institute of Physics, \'{E}cole Polytechnique F\'{e}d\'{e}rale de Lausanne (EPFL), 1015, Lausanne, Switzerland}
\author{G. Beaulieu}
\affiliation{Hybrid Quantum Circuits Laboratory (HQC), Institute of Physics, \'{E}cole Polytechnique F\'{e}d\'{e}rale de Lausanne (EPFL), 1015, Lausanne, Switzerland}
\affiliation{Center for Quantum Science and Engineering,\\ \ Institute of Physics, \'{E}cole Polytechnique F\'{e}d\'{e}rale de Lausanne (EPFL), 1015, Lausanne, Switzerland}
\author{O. Zilberberg}
\affiliation{Department of Physics, University of Konstanz, D-78457 Konstanz, Germany}
\author{P. Scarlino}
\affiliation{Hybrid Quantum Circuits Laboratory (HQC), Institute of Physics, \'{E}cole Polytechnique F\'{e}d\'{e}rale de Lausanne (EPFL), 1015, Lausanne, Switzerland}
\affiliation{Center for Quantum Science and Engineering,\\ \ Institute of Physics, \'{E}cole Polytechnique F\'{e}d\'{e}rale de Lausanne (EPFL), 1015, Lausanne, Switzerland}
\email{Second.Author@institution.edu}

\title{Band engineering and study of disorder using topology in compact high kinetic inductance cavity arrays}

\date{\today}

\begin{abstract}
Superconducting microwave metamaterials offer enormous potential for quantum optics and information science, enabling the development of advanced quantum technologies for sensing and amplification. 
In the context of circuit quantum electrodynamics, such metamaterials can be implemented as coupled cavity arrays (CCAs).
In the continuous effort to miniaturize quantum devices for increasing scalability, minimizing the footprint of CCAs while preserving low disorder becomes paramount.
In this work, we present a compact CCA architecture using superconducting NbN thin films manifesting high kinetic inductance. The latter enables high-impedance CCA ($\sim\SI{1.5}{\kilo\ohm}$), while reducing the resonator footprint.
We demonstrate its versatility and scalability by engineering one-dimensional CCAs with up to 100 resonators and with structures that exhibit multiple bandgaps.
Additionally, we quantitatively investigate disorder in the CCAs using symmetry-protected topological SSH modes, from which we extract a resonator frequency
scattering of $0.22^{+0.04}_{-0.03}\%$.
Our platform opens up exciting new prospects for analog quantum simulations of many-body physics with ultrastrongly coupled emitters.           

\end{abstract}

\maketitle

\begin{figure*}
    \includegraphics[width = \linewidth]{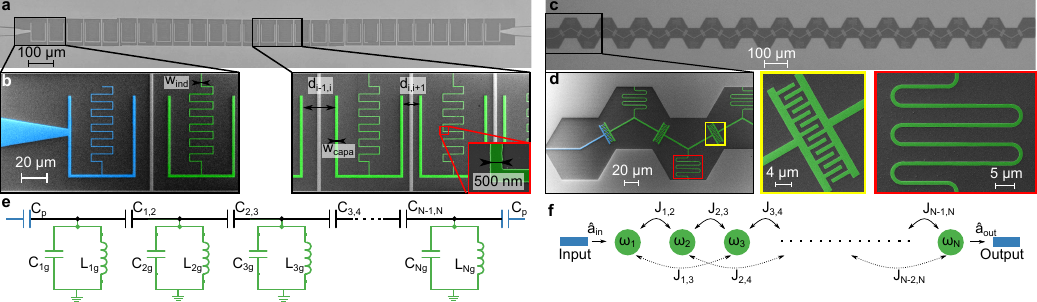}
    \caption{\label{fig:1}\textbf{High-impedance CCA.} \textbf{a}. Optical micrograph of a representative rectangular-design CCA comprising 25 resonators. The dark part is silicon while the light part is NbN. \textbf{b}. False-colored scanning electron micrographs of the zoomed-in regions of the CCA [marked by black frames in \textbf{a}]. The microwave port is colored in blue and the cavities in green. Inset: Further zoom-in on a portion of the inductor [marked by a red frame in \textbf{b}]. \textbf{c}. Optical micrograph of a representative hexagonal-design CCA comprising 26 resonators. \textbf{d}. False-colored scanning electron micrograph of the zoomed-in region of the CCA [marked in panel \textbf{c}]. The microwave port is colored in blue and the cavities in green. The insets [yellow (red) frame] show a zoom-in on the mutual capacitor between two cavities (the inductor of a cavity). \textbf{e}. Schematic of the lumped-element model of the CCAs, cf.~\eq\eqref{eq:Lk}. Each cavity is modeled as an LC resonator with an inductor $L_{ig}$ and a capacitor $C_{ig}$ to ground; the $i$ and $j$ cavities are coupled via a mutual capacitor $C_{i,j}$. The cavities at the edges of the CCA are coupled to the microwave ports via the capacitors $C_p$ in blue. \textbf{f}. Schematic of the corresponding chain Hamiltonian including first (second) neighbor interaction $J_{i,i+1}$ ($J_{i,i+2}$) between cavities $i$ and $i+1$ ($i+2$), cf.~\eq\eqref{eq:tight-bindingHamiltonian}. The input and output microwave ports are represented in blue, and $\hat{a}_{in(out)}$ indicates the input (output) field operator.}
\end{figure*}

Metamaterials made of superconducting circuits~\cite{carusotto2020photonic} have emerged as highly versatile platforms at the forefront of quantum technologies, offering a broad range of applications encompassing sensing~\cite{grimsmo2021QuantumMetamaterial}, amplification~\cite{macklin2015near}, and quantum information processing~\cite{ferreiraDeterministicGenerationMultidimensional2022,naik2017random}. 
Moreover, when quantum emitters are strongly or ultrastrongly coupled to microwave superconducting metamaterials that feature high quality and mode density ~\cite{puertasmartinezTunableJosephsonPlatform2019,kuzminSuperstrongCouplingCircuit2019,indrajeetCouplingSuperconductingQubit2020}, or to structured photonic baths~\cite{eggerMultimodeCircuitQuantum2013,hoodAtomAtomInteractions2016}, it provides a valuable framework for exploring many-body phenomena through analog quantum simulation~\cite{georgescuQuantumSimulation2014,altmanQuantumSimulatorsArchitectures2021,houck2012chip}.  
In particular, coupled cavity arrays (CCAs) have emerged as a flexible architecture for realizing artificial photonic materials in the tight-binding limit~\cite{saxena2023realizing,underwood2012low}, enabling the creation of band structures with varying complexity~\cite{kimAnalogQuantumSimulation2022,mostameEmulationComplexOpen2016}. 
These structures are even capable of realizing non-Euclidean geometries~\cite{kollarHyperbolicLatticesCircuit2019}, non-trivial tolopogical lattices~\cite{kimQuantumElectrodynamicsTopological2021,saxena2022photonic,owens2022chiral}, and flat bands~\cite{goblot2019nonlinear}, offering insights into complex many-body physics~\cite{zhangSuperconductingQuantumSimulator2023} and enabling various quantum information processing tasks~\cite{ ferreiraDeterministicGenerationMultidimensional2022,scigliuzzoControllingAtomPhotonBound2022,naik2017random}.

Conventional on-chip superconducting CCAs are realized with distributed coplanar waveguide (CPW)~\cite{gopplCoplanarWaveguideResonators2008} or lumped element LC~\cite{pozarMicrowaveEngineering2011} resonators, typically made of aluminum (Al) or niobium (Nb) superconducting thin films.
This architecture allows for arbitrary band engineering by tailoring the frequency and coupling of the cavities~\cite{kollarHyperbolicLatticesCircuit2019,morvanObservationTopologicalValley2021a,carusotto2020photonic}.
Despite this flexibility, the large physical footprint of a single resonator at frequencies $\sim$ \SI{5}{\giga\hertz} covers several millimeters' square~\cite{gopplCoplanarWaveguideResonators2008,liuQuantumElectrodynamicsPhotonic2017}, and can restrict the scalability of the array.
Recent research has explored alternative approaches, such as the replacement of the geometric inductance with the Josephson inductance of compact junction arrays~\cite{masluk2012Microwave,scigliuzzoControllingAtomPhotonBound2022,osman2023mitigation}. 
However, achieving control of the Josephson junction inductance with an imprecision below a few percent~\cite{moskalev2023optimization,osman_mitigation_2023}, remains challenging, leading to significant variability in cavity parameters and impacting the spectral properties of the CCAs.
Additionally, this approach introduces significant nonlinearity, affecting the higher excitation manifold of the CCA.
Despite the variety of approaches, the challenge of realizing a CCA made of unit cells that combine simultaneously high-quality, ultra compactness, and weak nonlinearities, while maintaining a low overhead in fabrication, still remains elusive. 
To fully harness the potential of CCAs while dramatically reducing their size, it is crucial to maintain low scattering of the cavity frequencies and inter-site coupling, as well as to develop methods to efficiently quantify the impact of disorder. 

In this work, we report on a compact and versatile lumped-element CCA architecture characterized by low disorder, with only $0.22^{+0.04}_{-0.03} \%$ deviation in resonator frequency. 
The resonators are made of high kinetic inductance NbN thin film resulting in compact inductors~\cite{samkharadzeHighKineticInductanceSuperconductingNanowire2016,niepceHighKineticInductance2019,frascaNbNFilmsHigh2023}. 
We show the versatility and scalability of the platform by engineering one-dimensional CCAs with up to 100 resonators with multiple band-structures. 
To efficiently quantify the amount of disorder in the system, we develop a \textit{topology-inspired} metric for assessing the resonators' frequency scattering by systematically exploring the in-gap mode distribution of CCAs that realize the Su-Schrieffer-Heeger (SSH) chain~\cite{suSolitonExcitationsPolyacetylene1980}. 
Due to the bulk-edge correspondence~\cite{asbothShortCourseTopological2016} in-gap modes are also sensitive to chiral symmetry-breaking disorder in the bulk of the CCA.
Thus, by focusing solely on the boundary modes we can infer the overall disorder present in the entire system.
Notably, the high kinetic inductance of our devices enables the realization of high-impedance resonators in the array.
This characteristic increases the coupling to charge degree of freedom of both superconducting~\cite{devoretCircuitQEDHowStrong2007} and semiconducting~\cite{stockklauserStrongCouplingCavity2017} qubits, enhancing the possibility to achieve the ultra-strong coupling regime~\cite{friskkockumUltrastrongCouplingLight2019}. 
This development lays the groundwork for integrating quantum emitters into our bath-engineered CCA environment.

\section{Platform}

We design, simulate, fabricate, and investigate 1-dimensional (1D) CCAs comprising rectangular  (\figs\ref{fig:1}\textbf{a} and \textbf{b}) and hexagonal geometries (\figs\ref{fig:1}\textbf{c} and \textbf{d}).
Each CCA is fabricated from a high kinetic inductance NbN thin film (see Methods) and can be modelled as an array of $N$ superconducting lumped-element LC resonators, as schematically represented in \fig\ref{fig:1}\textbf{e}. 
Each resonator is defined by a capacitor with total capacitance $C_{\Sigma,i} \approx C_{ig} + C_{i-1,i} + C_{i,i+1}$, where 
$C_{ig}$ represents the capacitance to ground of the $i^\text{th}$ cavity, which also shares mutual capacitances, $C_{i-1,i}$ and $C_{i,i+1}$ with its neighboring resonators.
These latter two capacitances can be adjusted by varying the spacing between the resonators $d_{i-1,i}$ and $d_{i,i+1}$, or by adjusting the interdigitated capacitor, see \figs\ref{fig:1}\textbf{b} and \textbf{d}.
The resonance frequency of the $i^\text{th}$ resonator, denoted as $\omega_i/2\pi$, is determined by  $1/\sqrt{L_{ig}C_{\Sigma,i}}$. 
Both $C_{ig}$ and $C_{i\pm1,i}$ can be tailored independently, while keeping $C_{\Sigma,i}$ and, consequently, the resonance frequency $\omega_i/2\pi$, constant. 
The inductance to ground, $L_{ig}$, of the nanowire inductor with width $w_{\text{ind}}$ and length $l_\text{ind}$ (\figs\ref{fig:1}\textbf{b}, \textbf{d}) can be expressed as~\cite{samkharadzeHighKineticInductanceSuperconductingNanowire2016,niepceHighKineticInductance2019}:
\begin{equation}
L_{ig} =  L_{\text{k}} + L_\text{geo}= L_{\text{k},\square}\frac{l_\text{ind}}{w_\text{ind}} + L_\text{geo},
\label{eq:Lk}
\end{equation}
with $L_{\text{k}}$  ($ L_{\text{k},\square}$) representing the (sheet) kinetic inductance,  and $L_\text{geo}$ accounting for the geometric inductance. 
In our case, the ratio $L_{\text{k}} / L_\text{geo} \sim 250$ indicates that the inductance of the CCA is completely dominated by its kinetic contribution.
Leveraging this property, we are able to engineer resonators with a significantly reduced footprint typically down to $50\times\SI{75}{\micro\metre\squared}$~\cite{frascaNbNFilmsHigh2023}. 
This size is remarkably smaller if compared to conventional lumped-element resonators ($\sim100\times$ smaller)~\cite{kimQuantumElectrodynamicsTopological2021,wangModeStructureSuperconducting2019} and CPW distributed resonators ($\sim 1500 \times$ smaller)~\cite{gopplCoplanarWaveguideResonators2008,kollarHyperbolicLatticesCircuit2019}.
Due to the large kinetic inductance, the resonators in the array present an impedance  $Z_i=\sqrt{L_{ig}/C_{\Sigma,i}}$ of approximately $\SI{1.5}{\kilo\ohm}$ ($\SI{0.8}{\kilo\ohm}$) for the rectangular (hexagonal) geometry. 
This high impedance enhances the capacitive couplings between resonators ($J_{i,i+1} \propto \sqrt{Z_iZ_{i+1}}$) and to quantum emitters ($g_i \propto \sqrt{Z_i}$).

The Hamiltonian of the system is derived following standard Lagrangian circuit quantization~\cite{vool2017introduction} (see Methods), and takes the form,
\begin{equation}
\begin{split}
     \hat{H} = &\sum_{i=1}^{N} \omega_i\hat{a}_i^\dagger\hat{a}_{i} + \sum_{q=1}^Q\sum_{i=1}^{N-q} J_{i,i+q}\left(\hat{a}_i^\dagger\hat{a}_{i+q} + h.c.\right)\,,
\end{split}
\label{eq:tight-bindingHamiltonian}
\end{equation}
up to the $Q^\text{th}$ order in coupling in $\hbar=1$ units.
Here, $\hat{a}_i$ and $\hat{a}_i^\dagger$ are the photonic annihilation and creation operators at site $i$.
The nearest-neighbor coupling terms, $J_{i,i+1}$, originate mainly from direct capacitive coupling. In our model, we neglect inductive coupling due to the resonators' high impedance, as it scales as $1/\sqrt{Z_i}$.
Coupling terms of order $q>1$ arise from two origins: stray capacitive coupling between next nearest-neighbor resonators and from the inversion of the capacitance matrix in the circuit's Lagrangian, the latter increases with the ratio $C_{i,i+1}/C_{\Sigma,i}$~\cite{supp} (see Methods). 
In the following, we focus on CCAs realized with resonators with degenerate frequencies, $\omega_r/2\pi$.
To ensure this degeneracy, we introduce ghost ports (see \fig\ref{fig:1}\textbf{b}) which guarantee the uniformity of the capacitive environment for both edge and bulk resonators.

\section{Band engineering \label{sec:bandEngineering}}
\begin{figure*}[t]
    \centering
    \includegraphics[width = 0.99\linewidth]{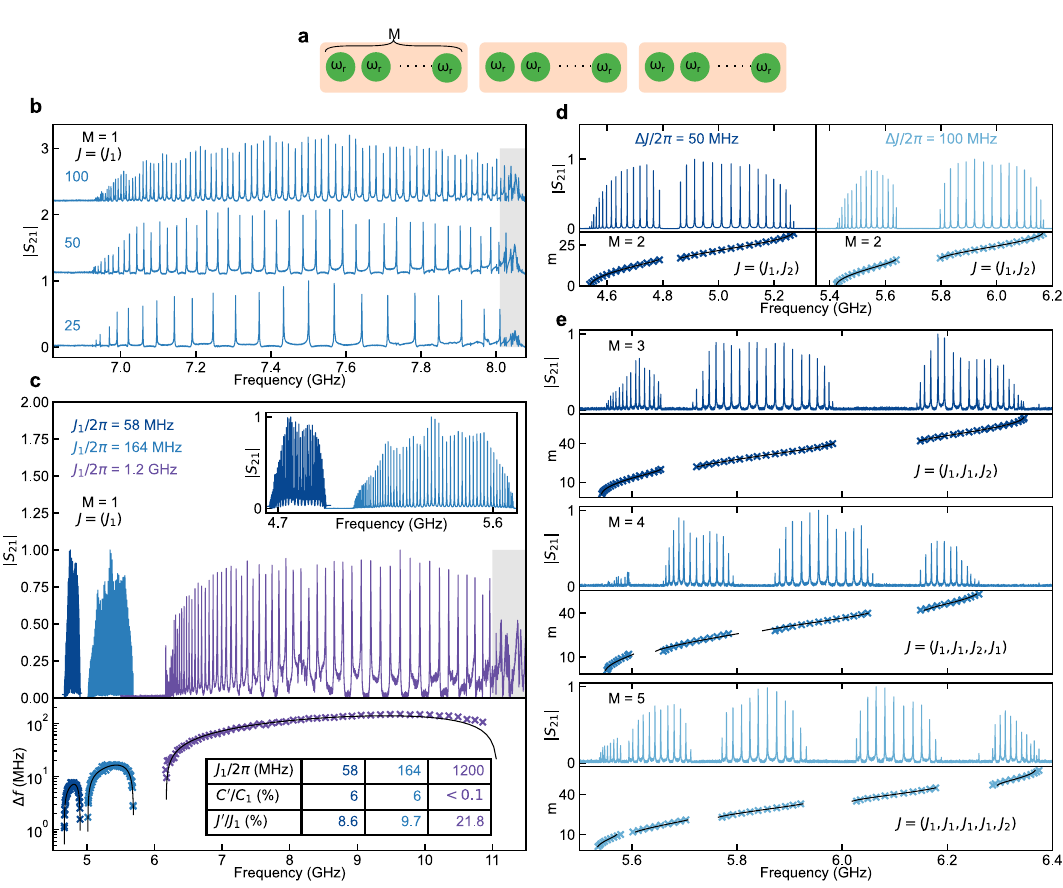}
    \caption{\textbf{Band engineering.}  \textbf{a}. General CCA schematic displaying the cavities in green, with a unit cell including $M$ cavities. \textbf{b}. Transmission spectrum $|S_{21}|$ for CCAs with $N = 25$, $50$ and $100$ cavities with rectangular design with $M = 1$. The shaded region around \SI{8.05}{\giga\hertz} highlights the presence of microwave chip slot modes. \textbf{c}. Top: Same as \textbf{b} for $M = 1$ CCAs with $J_1/2\pi = \SI{58}{\mega\hertz}$ (rectangular design), $J_1/2\pi = \SI{200}{\mega\hertz}$ (rectangular design), and $J_1/2\pi = \SI{1200}{\mega\hertz}$ (hexagonal design). Additional slot modes are visible around $\SI{11.1}{\giga\hertz}$. Inset: a zoom-in on the two lower coupling CCAs in the main panel. Bottom: Frequency difference, $\Delta f = f_i-f_{i+1}$, between two consecutive modes for the corresponding CCAs. The crosses represent the $\Delta f$ extracted for the CCAs, each plotted in relative to the averaged modes' frequencies $(f_i + f_{i+1})/2$. The continuous black lines indicate fits of the extracted $\Delta f$ according to the eigenmodes of the CCAs Hamiltonian (see Methods). Inset table collects the coupling $J_1/2\pi$, the stray capacitance ratio $C^\prime/C_\Sigma$, where $C^\prime = C_{i,i+2}$, and next nearest-neighbor couplings ratio $J^\prime/J_1$ for the three CCAs. \textbf{d}. Top: Same as \textbf{c} for dimerized CCAs (rectangular design, $M = 2$) with $\Delta J/2\pi = |J_2 - J_1|/2\pi \sim$ \SI{50}{\mega\hertz} (\SI{100}{\mega\hertz}) on the left (right). Bottom: Extracted modes (crosses) and their fits to the eigenmodes of the CCAs Hamiltonian (continuous line) (see Methods). \textbf{e}. Same as \textbf{c} for (top to bottom) unit cells with $M = 3$, $4$ and $5$.  All the transmission spectra are normalized by their maximum amplitude.}
    \label{fig:2}
\end{figure*}

We begin by characterizing the extensibility of the introduced compact CCA platform by measuring  chains with homogeneous coupling, i.e., $J_{i, i+1}=J_1$  and $J_{i, i+2}=J^\prime$, cf.~\fig\ref{fig:1}\textbf{d}. To this end, we set $C_{ig} = C_g$, $L_{ig} = L_g$ and $C_{i,i+1} = C_1$. 
Note that we have $M=1$ resonators per unit cell [cf.~\fig\ref{fig:2}\textbf{a}], and for $J_1 \gg J^{\prime}$, we expect a finite-size sampling of a cosine dispersion, i.e., the emergence of a passband centered around $\omega_r/2\pi$ with a span of $4J_1$~\cite{scigliuzzoControllingAtomPhotonBound2022}. 
In \fig\ref{fig:2}\textbf{b}, we report on the transmission spectra of several such homogeneous CCAs with $N$ up to $100$ sites. 
The transmission, $\left|S_{21}\right|$,  is measured in a cryogenic setup at $\SI{10}{\milli\kelvin}$ with a vector network analyzer~\cite{supp}. 
Each CCA transmission is normalized with respect to its maximum transmission amplitude. 
We observe $N$ distinct peaks, corresponding to the eigenmodes of the CCA. The modes at the center of the band have respectively larger coupling to the ports [higher peaks], $\kappa_\text{ext}$, and sparser frequency spacing, $\Delta f$, relative to the smaller coupling [lower peaks] and higher mode density at the band edges~\cite{scigliuzzoControllingAtomPhotonBound2022} (\fig\ref{fig:2}\textbf{b} and \cite{supp}). 
In all examined CCAs, we resolve the majority of the modes; for example, we detect as many as $\sim$ 90 distinct modes in the case of $N=100$ CCA. 
The missing modes can be attributed to two dominant factors:
(i) the modes at the edges of the pass-band have lower visibility, and (ii) to avoid erroneous counting, we exclude the frequency region around \SI{8.1}{\giga\hertz} (\fig\ref{fig:2}\textbf{b}), where chip slot modes are present.
On average, we extract individual mode single-photon internal dissipation rates $\kappa_\text{int}^\text{Mode}/2\pi$ of $\SI{100}{\kilo\hertz}$ for $\omega_r/2\pi = \SI{5}{\giga\hertz}$ ($Q_\text{int}^\text{Mode}\sim 50\times10^3$), indicating low-loss CCAs~\cite{supp}.
Remarkably, fabricating CCAs with a high number of cavities doesn't degrade the low-power quality factor of the device~\cite{frascaNbNFilmsHigh2023}.

We proceed to demonstrate high control over the inter-site coupling, see \fig\ref{fig:2}\textbf{c}.  
By redistributing the components contributing to the total capacitance, $C_\Sigma$, of each resonator, specifically adjusting the capacitance to ground, $C_g$, and mutual capacitance, $C_1$, we can modulate the inter-site coupling rate.
This allows us to engineer multimode environments with bandwidths ranging from approximately $\SI{230}{\mega\hertz}$ up to $\SI{4.8}{\giga\hertz}$, resulting in a free spectral range spanning from hundreds of MHz down to 1 MHz.
Considering the low-dissipation rates and the possibility to engineer narrow free spectral range, this architecture opens exciting prospects for exploring superstrong light-matter coupling~\cite{kuzminSuperstrongCouplingCircuit2019} and many-body Hamiltonians in the strongly non-linear regime~\cite{leger2019Observation,mehtaDownconversionSinglePhoton2023}.
Notably, we demonstrate exceptional control over high-quality CCAs comprising up to 100 resonators with a density of 5 resonators per \SI{100}{\micro\metre}, highlighting our capability to finely engineer the environment bandwidth.

To demonstrate a multiband spectrum, we consider configurations with $M$ up to 5 cavities per unit cell (\fig\ref{fig:2}\textbf{d} and \textbf{e}). 
We denote with $J_i$, where $i=1\ldots M-1$, the coupling between cavities within a unit cell (intracell coupling) and with $J_{M+1}$ the coupling between unit cells (intercell coupling).  
As we increase the number of elements in the unit cell, additional bands appear in the array spectrum~\cite{dana1985quantised}.
As such, bandgaps are expected to emerge in the midst of the CCA's spectrum, with up to $M$ passbands.  
In the dimer case ($M=2$), each resonator presents the same two coupling capacitances, in an alternating fashion, which automatically satisfies the resonant condition. 
However, for $M>2$, accomplishing the resonant frequency condition for all cavities requires precise control over the mutual and ground capacitances~\cite{supp}.

In all measurements in \fig\ref{fig:2}, the influence of higher-order coupling terms, $J^\prime$, are present and manifest in two primary aspects:
first, the mode distribution in the passband is asymmetric with respect to $\omega_r$, resulting in higher mode density at lower frequencies~\cite{supp}. 
Second, the mode coupling to the ports, $\kappa_\text{ext}$, for the low frequency eigenmodes is lower than for their higher frequency counterpart~\cite{supp}.
By fitting the CCA spectra (see Methods), we estimate $J^\prime\approx10\% \overline{J}$ for the rectangular designs (dominated by direct stray capacitive coupling), while for the hexagonal one, $J^\prime\approx20\% \overline{J}$ (due to high $C_{i,i+1}/C_{\Sigma,i}$ ratio), where $ \overline{J}$ is the mean of the nearest neighbor couplings in the CCA (see Table in \fig\ref{fig:2}\textbf{c}).
Furthermore, the asymmetry observed in the size of the bandgaps (\fig\ref{fig:2}\textbf{e}) can be attributed to systematic design imperfection~\cite{supp}.

\begin{figure*}[t]
    \centering
    \includegraphics[width=\linewidth]{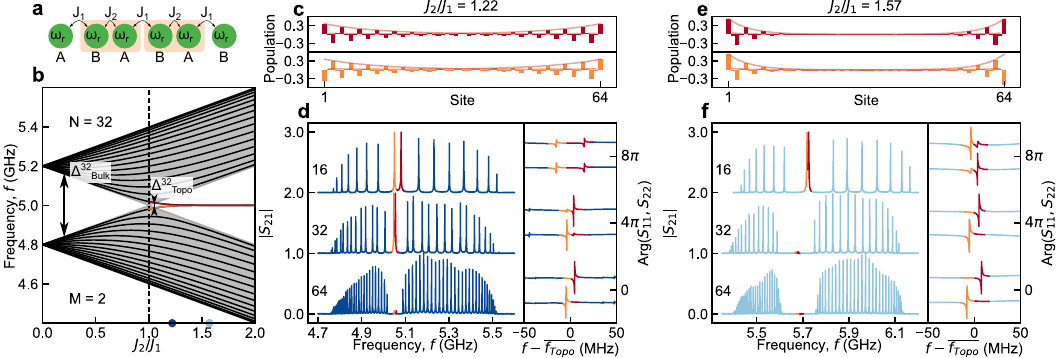}
    \caption{\textbf{Engineering SSH modes.} \textbf{a} Schematic of the SSH model. Each unit cell contains two cavities A and B, both with frequency $\omega_r$. $J_1$ and $J_2$ are respectively the intracell and intercell coupling. \textbf{b} Simulated phase transition diagram of the SSH model from trivial ($J_1>J_2$) to topological ($J_1<J_2$) phase, with $J^\prime = 0$. The black lines represent the bulk modes for a CCA with $N = 32$. For $J_1\neq J_2$, the system presents a bandgap of size $\Delta_{\text{Bulk}}^{32}$. In the non-trivial phase, two hybridized SSH modes (red and orange) are enabled at the center of the bandgap and are separated by $\Delta_{\text{Topo}}^{32}$. The grey area represents the phase transition diagram $N\rightarrow \infty$. 
    \textbf{c} (\textbf{e}) Simulated photonic population of the CCA with $N = 64$ in correspondence of the symmetric (red) and antisymmetric (orange) hybridized SSH modes in the weakly localized configuration, $J_2/J_1 = 1.22$ (strongly localized configuration, $J_2/J_1 = 1.57$) according to the eigenvectors of the CCAs Hamiltonian (see Methods). \textbf{d} (\textbf{f}) Left: Transmission spectrum $|S_{21}|$, for CCAs with $J_2/J_1 = 1.22$ ($J_2/J_1 = 1.57$) and $N = 16$, $32$ and $64$. Right: Reflection spectra Arg $S_{11,22}$, as a function of the frequency detuning $f-\overline{f_{Topo}}$, for a frequency region of 100 MHz around the SSH modes. $\overline{f_\text{Topo}}$ is the mean frequency of the two SSH modes. The modes in red and orange represent the symmetric and antisymmetric hybridized SSH modes, respectively. \label{fig:3}}
\end{figure*}

\section{Engineering localized modes}

Using our architecture, we demonstrated excellent control over the bulk spectrum of the CCA. Relying on topology, the creation of low-dimensional bound modes can prove useful for coupling to quantum emitters~\cite{Jackiw1976,Teo2010,gao2020dirac,kovsata2021second}. 
Remaining in a 1D chain geometry, we turn to engineer CCAs in the topologically non-trivial SSH configuration (\fig\ref{fig:3})~\cite{suSolitonExcitationsPolyacetylene1980,asbothShortCourseTopological2016}.
The SSH chain is a ubiquitous chiral symmetry protected topological phase of matter that manifest in 1D systems. Its bands exhibit a  quantized bulk polarization, with associated mid-gap 0D boundary modes.
The model has been extensively studied in photonic CCAs~\cite{saxenaPhotonicTopologicalBaths2022}, cold-atoms~\cite{atalaDirectMeasurementZak2013,lohse2016thouless}, polaritonics~\cite{st-jeanLasingTopologicalEdge2017}, and optomechanical arrays~\cite{youssefiTopologicalLatticesRealized2022}, and used to engineer directional topological waveguide QED~\cite{bello2019unconventional,kimQuantumElectrodynamicsTopological2021}.

Our microwave photonic analogue of the SSH model comprises a dimerized ($M=2$) chain  (\fig\ref{fig:3}\textbf{a}).
The intra- ($J_1$) and inter-cell ($J_2$) hopping are alternating, leading to a two bulk-band spectrum. However, a gap closing occurs when $J_1 = J_2$ (\fig\ref{fig:3}\textbf{b}). The gap closing marks a topological phase transition between the topologically-trivial  ($J_2<J_1$, cf.~\fig\ref{fig:2}\textbf{d})
and the topological nontrivial cases ($J_1<J_2$, cf.~\fig\ref{fig:3}). A quantized jump in the bulk polarization of the chain distinguishes between the two cases, where in the latter it implies the appearance of two degenerate mid-gap boundary modes.  

The Hamiltonian of our SSH CCA, in quasimomentum space reads~\cite{perez-gonzalezInterplayLongrangeHopping2019,supp}:
\begin{equation}
\begin{split}
    \hat{H}(k) =&\left(\omega_0 + 2J^\prime\cos{kd}\right)\tau_0\\
    &+\left(J_1 + J_2e^{-jkd}\right)\tau_x+\xi\,\tau_z
\end{split}
\label{eq:k-spaceHam}
\end{equation}
where $k$ is the reciprocal wavevector, $d$ is the lattice constant (distance between the unit cells), and $\tau_0$, $\tau_x$ and $\tau_z$ represent the Pauli matrices. Note that the appearance of next nearest-neighbor hopping $J^\prime$, and disorder terms $\xi$ due to fabrication imperfections lead to deviation from the standard SSH model~\cite{perez-gonzalezInterplayLongrangeHopping2019,supp}. The latter breaks the chiral symmetry and will move the topological boundary modes away from the middle of the gap. Disorder in $J_i$ does not break the chiral symmetry and will bear a lesser impact on the boundary modes~\cite{asbothShortCourseTopological2016,supp}.

In a finite-size CCA (\fig\ref{fig:3}\textbf{a}), the tails of the mid-gap boundary states overlap in the bulk, resulting into a finite hybridization that gives rise to a frequency splitting according to
\begin{equation}
     \Delta_\text{Topo}^N/2 \propto e^{-((N-1)/\zeta)},
     \label{eq:deltaTopo}
\end{equation}
where $\zeta$ is the boundary modes' spatial localization
\begin{equation}
    \zeta = \frac{1}{\log{J_2/J_1}}\,.
    \label{eq:zeta}
\end{equation}
The degree of hybridization between the SSH states depends on the chain's size, $N$, and the coupling ratio, $J_2/J_1$.
Correspondingly, the hybridized SSH states form symmetric and antisymmetric superpositions between the left and right boundary states, see \figs\ref{fig:3}\textbf{c} and \textbf{e}.
In the remainder of the manuscript, we will refer to the \textit{hybridized SSH states} as \textit{SSH modes}.

We experimentally investigate two distinct configurations: a weakly-localized case with $J_2/J_1 = 1.22$ (\figs\ref{fig:3}\textbf{c, d}) and a strongly-localized case with $J_2/J_1 = 1.57$ (\figs\ref{fig:3}\textbf{e, f}). 
These configurations present CCAs with $N = 16$, $32$, and $64$ resonators.
Measurements of CCA transmission, $S_{21}$, reveal a significant reduction in the amplitude of the SSH modes as the size of the CCA increases. 
This reduction can be attributed to the decreasing overlap of the localized edge states in the bulk region, resulting in reduced coupling and, therefore, reduced transmission between the two microwave ports~\cite{supp}.
This trend is also visible in the behavior of the phase shift of the SSH modes measured in reflection ($S_{11}$, $S_{22}$)~\cite{meidan2011topological} (right panel of \figs\ref{fig:3}\textbf{d} and \textbf{f}).
As the modes' hybridization reduces, the phase shift of the SSH modes becomes more prominent, indicating a stronger coupling to the microwave ports due to localization at the boundary.

As for the topologically trivial CCAs (\sect\ref{sec:bandEngineering}), next nearest neighbor coupling have an influence on the SSH-CCAs spectra.  
We expect that the two edge modes for $J^\prime \neq 0$ do not exhibit anymore perfect localization on a single sub-lattice of the unit cell~\cite{supp}. 
Instead, some photonic population extends into the neighboring sub-cell, thereby breaking chiral symmetry, even in the absence of a $\tau_z$ term in the Hamiltonian in \eq\eqref{eq:k-spaceHam}.
For $J^\prime$ smaller than the bandgap, the SSH modes retain partial protection~\cite{song2015quantization,kremerSquarerootTopologicalInsulator2020}.
$\Delta_\text{Topo}^N$ exponentially decreases as a function of $N$, even in the presence of non-zero $J^\prime$.

\section{Disorder \label{sec:disorder}}
\begin{figure*}
    \centering
    \includegraphics[width = \linewidth]{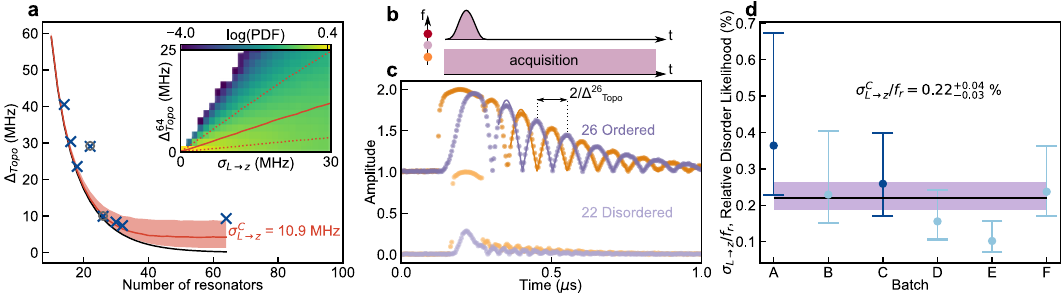}
    \caption{\textbf{Using topology to study disorder.} {\color{red} }\textbf{a.} Study of the frequency splitting, $\Delta_\text{Topo}^N$, between the hybridized SSH modes for CCAs with different $N$ ($J_2/J_1 = 1.22$). The cross and the circles show $\Delta_\text{Topo}^N$ extracted for CCAs with $N = $ 14, 16, 18, 22, 26, 30, 32 and 64, respectively from spectroscopy and time resolved measurements. The continuous black line represents the expected evolution of $\Delta_\text{Topo}^N$ \textit{vs} $N$ for the disorderless case according to the eigenmodes of the CCAs Hamiltonian (see Methods). The continuous red line represents the median of the most likely evolution (see panel \textbf{d}) of  $\Delta_\text{Topo}^N$ \textit{vs} $N$ in the presence of disorder with standard deviation $\sigma^C_{L\rightarrow z} = \SI{10.3}{\mega\hertz}$ applied to $\omega_r$ of the cavities in the CCAs. The shaded region represents the $1\sigma$ uncertainty on the estimation of $\Delta_\text{Topo}^N$. The inset shows the simulated logarithm of the Probability Density Function ($\text{PDF}$) of $\Delta_\text{Topo}^{64}$ as a function of $\sigma_{L\rightarrow z}$. The red line shows the median for each disorder realization. The red dotted lines show the $1\sigma$ standard deviation. \textbf{b.} Pulse sequence used for the time-resolved measurement. A gaussian pulse is sent at a frequency between the hybridized SSH modes from one of the edge of the CCA, while the signal is acquired on both side of the CCA. The orange (red) dot on the $y$-frequency axis on the left side, highlights the frequency of the antisymmetric (symmetric) hybridized SSH modes. \textbf{c.} Time traces of the transmitted $|S_{21}|$ and reflected $|S_{11}|$ signals of the CCAs with $N = 26$ and $22$ in panel (a). Each data points is averaged 20000 times. The continuous lines highlight the fit done with an exponentially decaying cosine~\cite{supp}. \textbf{d.} Disorder likelihood extracted for different batches of devices. The dots highlight the maximum likelihood extracted for each batch. The error bars show the respective FWHM. The black line shows the combined likelihood extracted among the different batches. The shadowed purple area shows its FWHM. The color code is according to \fig\ref{fig:3}\textbf{d} and \textbf{f}. Batch $A$ ($C$), contains the CCAs with $N = 16$, $32$ and $64$ ($14$, $18$, $22$, $26$, $30$) of panel \textbf{a}.\label{fig:4}}
\end{figure*}

We have demonstrated the scalability and versatility of the architecture, exhibiting a high degree of control and suggesting minimal resonant frequency scattering between the cavities.
However, some amount of disorder remains evident in the spectra of the CCAs.
In the bulk, disorder manifests in deviations from a smooth envelope profile of the mode's transmission and in displacement of mode frequencies from their expected dispersion relation (\figs\ref{fig:2} and \ref{fig:3}).
Quantifying disorder from the bulk modes is challenging due to the system complexity.
On the other hand, SSH modes, despite being localized at the edges of the CCA, are readily probed. Crucially, due to the bulk-edge correspondence~\cite{asbothShortCourseTopological2016} they are also sensitive to chiral symmetry-breaking disorder in the bulk of the CCA.
Hence, they can serve as a reliable indirect probe to quantitatively assess the extent of disorder in the bulk, solely by analysing the behavior of the two SSH modes.

To investigate the impact of disorder on the SSH modes, we fabricate and characterize additional SSH devices~\cite{supp} with a coupling ratio of $J_2/J_1 = 1.22$, as illustrated in \fig\ref{fig:3}\textbf{d}. 
In \fig\ref{fig:4}\textbf{a}, we present a dataset comprising $\Delta_\text{Topo}^N$ values extracted from the measured CCAs as a function of chain length $N$.
Notably, while $\Delta_\text{Topo}^N$ exhibits the expected exponential decay with respect to $N$, it does not asymptotically approach zero, demonstrating significant deviations from the theoretical prediction (black \textit{vs} red line \fig\ref{fig:4}\textbf{a}).

To rigorously account for this observation, we conduct numerical simulations that introduce Gaussian noise, denoted as $\sigma_L$, applied to the inductance values of all resonators within the chain. 
The choice to introduce scattering in $L_g$ as the main noise source is motivated by the fact that the inductors have the most critical dimension in the resonator design, rendering them more susceptible to scattering during the fabrication process.
The $\sigma_L$ noise applied to the inductors induces both $\tau_z$ and $\tau_x$ type of disorder in \eq\eqref{eq:k-spaceHam}, impacting respectively the resonant frequency and the coupling of the resonators in the CCA. 
Our analysis primarily focuses on the principal effect of $\sigma_L$ scattering namely $\tau_z$-type disorder (breaking chiral symmetry), which we refer to as $\sigma_{L\rightarrow z}$.

In the inset of \fig\ref{fig:4}\textbf{a}, we present the simulated Probability Density Function ($\text{PDF}$) for $\Delta_\text{Topo}^{64}$ as a function of $\sigma_{L\rightarrow z}$ (see Methods).
For $\sigma_L = 0$, $\Delta_\text{Topo}^{64} \sim 0$, with $J_2/J_1 = 1.22$.
As the disorder increases, the probability of observing $\Delta_\text{Topo}^{64}$ values higher than $\sim 0$ also increases, along with the standard deviation.

In the measurements shown in \fig\ref{fig:4}\textbf{a}, a notable deviation is observed for $\Delta_\text{Topo}^{22}$ compared with the general trend.
To gain further insight into the source of this deviation, we perform time-resolved measurements of the SSH modes amplitude, using the pulse sequence illustrated in \fig\ref{fig:4}\textbf{b}.
This sequence involves sending a Gaussian pulse at a frequency between the two SSH modes, leading to beating oscillations in time between the modes' population if the two modes are hybridized, with a frequency of the beating corresponding to the coupling rate $\Delta_\text{Topo}^{N}/2$~\cite{supp}. 
The results of this measurement are presented in \fig\ref{fig:4}\textbf{c} for the cases corresponding to the two circled data points in \fig\ref{fig:4}\textbf{a}: $N = 22$ (the outlier point in \fig\ref{fig:4}\textbf{a}) and $N = 26$ (a representative of devices following the trend in \fig\ref{fig:4}\textbf{a}).
For the $N = 26$ SSH-CCA, we distinctly observe beatings between the two SSH modes at a frequency of approximately $4.97$ MHz, aligning with the value of $\Delta_\text{Topo}^{26} = \SI{4.96}{\mega\hertz}$ extracted from spectroscopy measurements (\fig\ref{fig:4}\textbf{a}).
In contrast, for the $N = 22$ CCA, we observe a significantly reduced visibility of the beating pattern, indicating weak coupling to the edge microwave ports and, therefore, suggesting that the modes are localized not at the edge but more in the bulk of the CCA.
This could be due to two effects: a strong disorder at the edge lifts the resonant frequency of a resonator or a strong impurity in the bulk (strongly detuned resonator) effectively divides the chain and quenches the transmission.
We utilize these time-resolved measurements to identify devices with strong local disorder that no longer conform to our Gaussian disorder model of the SSH.

To evaluate the overall disorder introduced during the fabrication process, we conduct a comprehensive statistical analysis involving 26 CCAs in the SSH configuration. 
These 26 devices, manufactured during different fabrication runs, have been designed in the two SSH configurations depicted in \fig\ref{fig:3}~\cite{supp}.
To quantify the disorder, we extract $\Delta_\text{Topo}^N$ from spectroscopy measurements for all the tested devices and generate their associated $\text{PDF}$s as functions of $\sigma_L$ and $N$ for each sample batch $S$. 
For each batch, we compute the likelihood of the inferred frequency fluctuation with respect to the model as a function of $\sigma_L$, which we refer to as likelihood function, defined as follows:
\begin{equation}
    \mathcal{L}_{S}(\sigma_L) = \prod_{i}\text{PDF}_S(N_i,\Delta_{\text{Topo},i}^{N_i};\sigma_L),
\end{equation}
where $N_i$ and $\Delta_{\text{Topo},i}^{N_i}$ represent respectively the number of resonators in the CCA and the SSH mode frequency splitting of the $i^\text{th}$ CCA. 
We report the extracted maximum likelihoods and their full width at half maximum (FWHM) in \fig\ref{fig:4}\textbf{d}.
To obtain an estimate of the disorder across all the tested batches, we computed the combined likelihood, represented by the black line in the same figure (see Methods).
Our analysis yield an extracted relative disorder value of $\sigma_{L\rightarrow z}^C/f_r = 0.22_{-0.03}^{+0.04}\%$, equivalent to an absolute disorder value of $\sigma_{L\rightarrow z}^C = 10.97_{-1.59}^{+2.28}$ MHz for $\omega_r/2\pi\sim\SI{5}{\giga\hertz}$.
This represents a minimal frequency scattering, especially considering the high compactness of the implemented CCAs, and it is comparable to what is achieved with lattices of CPW resonators (but with $10^{2-3}$ larger footprint)~\cite{underwood2012low} and state of the art frequency scattering control of advanced MKIDs detector arrays ~\cite{mckenneyTileandtrimMicroresonatorArray2019,liReducingFrequencyScatter2022}.

\section{Conclusions}

We have introduced a novel platform based on coupled cavity arrays (CCAs) utilizing high kinetic inductance NbN thin films, which serve as compact multipurpose high-impedance metamaterials in the microwave domain. 
The compactness of each cavity allows for the integration of 1D CCAs with up to 100 resonators within a few millimeters of sample space.
The remarkable versatility of our CCA platform has been demonstrated through the creation of CCAs with bandwidths ranging from a few 100 MHz up to 4.5 GHz and the engineering of multiple bandgaps.
Importantly, all fabricated devices exhibited mode dispersion in excellent agreement with our exact models.
Furthermore, using the SSH chain's in-gap modes, we extracted a small resonator frequency scattering of $\sigma_{L\rightarrow z}^C/f_r = 0.22^{+0.04}_{-0.03}\%$.

Our findings pave the way for advancing technological applications and fundamental investigations using multimode light-matter systems.
This platform will allow for a straightforward extension to very large-scale 1D and 2D multimode systems with up to $10^4$ cavities on a single 5$\times$\SI{5}{\milli\metre\squared} chip.
The versatility of our platform in controlling mode densities presents exciting prospects for exploring devices where emitters are coupled to high-impedance multimode environments~\cite{kuzminSuperstrongCouplingCircuit2019}.
This provides the means to study the ultrastrong coupling limit in both the passband of the CCA ~\cite{legerRevealingFinitefrequencyResponse2023} and atom-photon bound states close to the band edges~\cite{scigliuzzoControllingAtomPhotonBound2022,vrajitoarea2024ultrastrong}.
This will allow to investigate quantum-impurity models like spin-boson~\cite{messingerLefthandedSuperlatticeMetamaterials2019} or Frenkel-Holstein~\cite{kimAnalogQuantumSimulation2022,mostameEmulationComplexOpen2016} type Hamiltonians.
Additionally, the compact nature of the resonators facilitates coupling to superconducting qubits at multiple points, potentially with non-trivial phase delays.
This makes our architecture a natural platform for studying giant-atom photon bound states~\cite{soro2022ChiralGiant} interacting with structured environments~\cite{soro2023Interactionbetween}.

On the other hand, photon lattices also offer promising avenues for future experiments aimed at investigating quantum phase transitions~\cite{fitzpatrick2017observation,carusotto2020photonic}. 
By effectively reducing random disorder, cavity arrays can be fabricated with controlled levels of disorder, potentially enabling the study of many-body localization effects~\cite{mehtaDownconversionSinglePhoton2023}. 
While our current work remains non-interacting, the incorporation of interactions is feasible through the inherent nonlinearity present in high kinetic inductance materials, resulting in both $\chi^2$ and $\chi^3$ nonlinearities~\cite{frasca2024three}, or by integrating qubits into each resonator~\cite{fitzpatrick2017observation}. 
We aim to leverage nonlinearities in CCAs as a novel tool for studying driven-dissipative phase transition~\cite{Soriente2021,Ferri2021,fitzpatrick2017observation}. 
Moreover, these lattices facilitate the creation of unique devices capable of hosting photons in curved spaces~\cite{kollarHyperbolicLatticesCircuit2019}, gapped flat band~\cite{leykam2018ArtificialFlat}, and novel forms of qubit-qubit interaction~\cite{kimQuantumElectrodynamicsTopological2021,pakkiam2023qubit}.

\begin{acknowledgments}
The authors thank Jan Ko{\v{s}}ata and Andrea Bancora for stimulating discussions. 
The authors thank the EPFL physics mechanical workshop for help in setting up the lab. 
This research was partly supported by the Swiss National Science Foundation (SNSF) through the grants contract number 200021\_200418 and 206021\_205335. 
P.S. also acknowledge the Swiss State Secretariat for Education, Research and Innovation (SERI) through the grant contract number MB22.00081 and the NCCR SPIN, funded by SNSF. S.F. acknowledges the support of SNF Spark project 221051. M.S. acknowledges support from the EPFL Center for Quantum Science and Engineering postdoctoral fellowship. O.Z. acknowledges funding from the Deutsche Forschungsgemeinschaft (DFG)
via project number 449653034 and through SFB1432, as well as the Swiss National Science Foundation (SNSF) through the Sinergia Grant No.~CRSII5\_206008/1.\\

\textbf{Contributions :}
V.J., S.F., V.J.W. and P.S. designed the experiment. V.J., S.F., F.O., F.D.P. and D.S. fabricated the devices. S.F. and F.O. developed and optimized the fabrication recipes. S.F. developed and optimized the NbN deposition. V.J. performed the measurements. V.J. analyzed the data with supervision from L.P. and inputs from G.B.. V.J. numerically simulated the model with supervision from M.S.. P.S. and O.Z. supervised the experimental and theoretical parts of the project. V.J., S.F., L.P., M.S., O.Z. and P.S. contributed to the writing of the paper.

\end{acknowledgments}

\section{\label{sec:methods} Methods}

\paragraph*{\textbf{Fabrication: }}

The fabrication recipe is detailed in~\cite{frascaNbNFilmsHigh2023}.
We fabricate planar coupled cavity arrays (CCAs) based on lumped LC resonators by etching 13~nm-thick NbN film, with typical sheet kinetic inductance $L_{\text{k},\square}$ of 100~pH/$\square$. 
The fabrication process commences with a 2-minute immersion in a 40$\%$ HF bath to eliminate the native oxide layer and potential surface impurities from a 100 mm silicon wafer, which is of high-resistivity ($\ge 10~k\Omega$cm) and has a $\braket{100}$ orientation.
Using a Kenosistec RF sputtering system at room temperature, NbN films are bias sputtered following the method described in~\cite{frascaNbNFilmsHigh2023,daneBiasSputteredNbN2017} with Ar/N$_2$ flows of 80/7 sccm respectively and a deposition pressure of \SI{5}{\micro\bar}. 
Optical lift-off technique is employed to deposit Ti/Pt alignment markers, followed by a dehydration step at 150\degree C for 5 minutes. 
80~nm-thick CSAR positive e-beam resist is then spun on the wafer, which is subsequently baked at 150\degree C for 5 minutes.
Employing electron beam lithography (Raith EBPG5000+ at 100 keV), the resist is patterned to form the desired devices. 
This is achieved by developing the resist in amyl acetate for 1 minute, followed by rinsing in a 9:1 MiBK:IPA solution. 
In order to transfer the pattern onto the NbN, a reactive ion etching process is employed using a CF$_\text{4}$/Ar mixture. 
The etching is carried out with a power of 15 W, using a stepped approach consisting of 10 steps, each lasting for 1 minute. 
These etching steps are alternated with 1-minute purges using Ar gas. 
This stepped etching technique has proven advantageous as it reduces the damage caused to the CSAR resist due the etching process, thereby facilitating its subsequent stripping without the need for plasma oxygen, which may damage the underlying superconductor. 
The resist is then stripped using Microposit remover 1165 heated to 70\degree C. 
Finally, the wafer is coated with a \SI{1.5}{\micro\metre} AZ ECI 3007 positive photolithography resist to protect the devices before being diced.
\\

\paragraph*{\textbf{Model: }}
In this section, we derive the Hamiltonian of the CCAs using standard circuit quantization~\cite{vool2017introduction}. 
We consider a chain composed of $N$ capacitively coupled LC resonators as depicted in \fig\ref{fig:1}. 
Each $i^{\text{th}}$ resonator possesses an inductance $L_{ig}$ connected to ground and capacitance $C_{ig}$ to ground. 
Resonators $i$ and $j$ are mutually coupled via the coupling capacitance $C_{i,j}$ between the two resonators. 
The potential energy in the inductors can be expressed as:
\begin{equation}
    E_L = \frac{1}{2}\sum_{n=1}^N\frac{\phi_n^2}{L_{ng}},
\end{equation}
where $\phi_n$ denotes the flux at node $n$. 
The total kinetic energy stored in the chain's capacitors is given by:
\begin{equation}
    E_C = \frac{1}{2}\left[\sum_{n=1}^N C_{ng}\dot{\phi}_n^2 + \sum_{i,j}C_{i,j}(\dot{\phi}_i-\dot{\phi}_j)^2\right],
\end{equation}
where $\dot{\phi}_n$ represents the electric potential at node $n$. 
We neglect mutual inductance-induced coupling due to the high impedance of the resonators~\cite{devoretCircuitQEDHowStrong2007}.
We can now write the Lagrangian, $\mathds{L}$, of the circuit as,
\begin{align}
    \mathds{L} &= E_C - E_L\\
    & = \frac{1}{2}\sum_{n=1}^N\left[C_{ng}\dot{\phi}_n^2 - \frac{\phi_n^2}{L_{ng}}\right] + \frac{1}{2}\sum_{i,j} C_{i,j}\left(\dot{\phi}_i^2-\dot{\phi}_j^2\right).
\end{align}
It can be written in a matrix form as
\begin{equation}
    \mathds{L} = \frac{1}{2}\dot{\bm{\phi}}^T\left[C\right]\dot{\bm{\phi}} - \frac{1}{2}\bm{\phi}^T\left[L^{-1}\right]\bm{\phi},
\end{equation}
with the vectors $\dot{\bm{\phi}}_n^T = (\dot{\phi}_1, \dot{\phi}_2,\hdots, \dot{\phi}_N)$ and $\bm{\phi}_n^T = (\phi_1, \phi_2,\hdots ,\phi_{N})$. 
The capacitance matrix is defined as,
\begin{equation}
    [C]_{ij} = 
    \begin{cases}
    C_{\Sigma i},\text{ if } i = j,\\
    -C_{i,j},\text{ if } i \neq j \text{ and } |i-j|\leq3,\\
    0,\text{ if } i \neq j \text{ and } |i-j|>3,
    \end{cases}
    \label{methods:eq:numCapa}
\end{equation}
where we only consider mutual capacitances where $|i-j|\leq3$. 
$C_{\Sigma i}$ is the total capacitance of the $i^\text{th}$ cavity defined as,
\begin{equation}
    C_{\Sigma i} = C_{ig} + \sum_{n\neq i}^N C_{i,n}.
\end{equation}
The inverse inductance matrix is defined as,
\begin{equation}
    [L^{-1}]_{ij} =
    \begin{cases}
    1/L_{ig},\text{ if } i = j,\\
    0,\text{ if } i \neq j.
    \end{cases}
\end{equation}
We now introduce the node charge variable canonically conjugated to the node flux $\bm{\phi}_n$
\begin{equation}
    \bm{Q} = \frac{\partial\mathds{L}}{\partial\dot{\bm{\phi}}},
\end{equation}
with $\bm{Q}^T = (Q_1, Q_2, \hdots, Q_N)$.

For the given system, the charge variables are $\bm{Q} = [\bm{C}]\dot{\bm{\phi}}$.
Making use of the matrix formalism, the CCA Hamiltonian $H$ then reads
\begin{equation}
    H = \frac{1}{2}\bm{Q}^T[C^{-1}]\bm{Q} + \frac{1}{2}\bm{\phi}^T[L^{-1}]\bm{\phi}.
    \label{methods:eq:HamLC}
\end{equation}
The real space Hamiltonian can be found to be~\cite{vool2017introduction}:
\begin{equation}
    \bm{H_n}/\hbar = \sqrt{[C^{-1}][L^{-1}]},
\end{equation}
and have the following matrix form
\begin{equation}
    \frac{\bm{H_n}}{\hbar} = \begin{pmatrix}
        \omega_1 & J_{1,2} & J_{1,3} & \hdots & \hdots & J_{1,N} \\
        J_{2,1} & \omega_2 & J_{2,3} & \ddots & \ddots & \vdots \\
        J_{3,1} & J_{3,2} & \omega_3 & \ddots & \ddots & \vdots\\
        \vdots &\ddots & \ddots & \ddots & \ddots & \vdots \\
        \vdots & \ddots & \ddots & \ddots & \omega_{N-1} & J_{N-1,N} \\
        J_{N,1} & \hdots & \hdots & \hdots & J_{N,N-1} & \omega_N
    \end{pmatrix}.
    \label{methods:hamMat}
\end{equation}
For small $C_{i,i+1}/C_{\Sigma,i}$ ratio, and without stray next nearest neighbor capacitances, the coupling is,
\begin{equation}
    J_{i,j} = \frac{\sqrt{\omega_i\omega_j}}{2}\frac{C_{i,j}}{\sqrt{C_{\Sigma,i}C_{\Sigma,j}}}.
\end{equation}
However, non-negligible additional contributions arise from both parasitic capacitances $C_{i,l\neq j}$ and large $C_{i,i+1}/C_{\Sigma,i}$ ratio, modifying the Hamiltonian~\cite{supp}.
In that case, one cannot resort to simple analytical formulas and a full numerical diagonalization is required.
Eigenvalues of \eq\eqref{methods:hamMat} represent the frequencies of the modes $\omega_k$ and eigenvectors represent the spatial localization of the modes. 
We utilize this model to fit the CCA modes' frequencies, as detailed in the following section.\\

\paragraph*{\textbf{Extraction of parameters: }}

The estimation of the parameters is performed by extracting the modes' frequencies from the device's spectrum and fitting them to the eigenvalues of \eq\eqref{methods:hamMat}.
The fitting process involves 5 to 8 fitting parameters, depending on the specific design, i.e. if the CCA is dimerized, trimerized, etc \dots. 
Initially, we make the assumption that each fitted CCA is disorder-free and uniform.
The influence of disorder is studied in the following section.
The fitting parameters include: 
\begin{itemize}
    \item $C_g$, the capacitance to ground.
    For a single resonator ($M=1$) or a dimer ($M=2$), the capacitive environment is automatically identical for each resonator and we use a single value for $C_g$.
    However, for $M>2$, the capacitive environment of each cavity is not identical~\cite{supp} and $C_g$ becomes a list, $\Vec{C_g}$, comprising the different $C_{ig}$ within a unit cell.
    \item $L_g$, the inductance to ground. 
    For $M\leq 2$, since the capacitive environment is similar for each cavity, having a constant inductance ensures a constant frequency for each cavity in the unit cell and we use a single value for $L_g$.
    For $M>2$, in order to keep the resonant frequency constant, $L_g$ is adjusted for each cavity~\cite{supp}.
    In this case, it becomes a list $\Vec{L_g}$.
    \item $\Vec{C_c}$, the coupling capacitances, which form a list increasing with the size of the unit cell, i.e. for $M=2$, $C_c = \left(C_1, C_2\right)$, $C_1$ and $C_2$ being the intra- and inter-cell capacitances.
    \item $C_{i,i+2}/\overline{C}$, the ratio of second neighbor coupling capacitances over $\overline{C}$, the mean value of $\Vec{C_c}$.
    \item $C_{i,i+3}/\overline{C}$, the ratio of third neighbor coupling capacitances over $\overline{C}$.
\end{itemize}

For the fits to converge, we must either fix $L_g$ or $C_g$ as they both contribute comparably to the resonant frequency of the cavities and the coupling between cavities.
To this end, we choose to fix $L_g$.
We determine the value of $L_g$ through finite-element microwave simulations, where we estimate the sheet kinetic inductance of the film using Sonnet simulation software. 
This process involves three steps:

1. We initially fit the modes of the measured spectrum with the eigenvalues of \eq\eqref{methods:hamMat}, fixing $L_g$ with an initial guess. 
This provides a precise estimation of the resonant frequency of the cavities but not of the other parameters.

2. We then conduct a simulation for a single cavity in Sonnet, while sweeping the kinetic inductance, $L_{\text{k},\square}$.
The simulated cavity has a capacitive environment equivalent to that of the cavities in the fitted CCA, ensuring accurate estimation of the resonant frequency.
Subsequently, we extract and fit the resonant frequency of the cavity as a function of the kinetic inductance using the following function for the frequency:
\begin{equation}
    f = \frac{1}{2\pi\sqrt{(L_{\text{k},\square}\frac{l_\text{ind}}{w_\text{ind}}+L_{\text{geo}})C_\Sigma}},
    \label{met:eq:fvsLsq}
\end{equation}
where $C_\Sigma$ represents the total capacitance of the resonator and $L_{\text{geo}}$ is the geometric inductance.
The extracted $L_{\text{geo}}$ is usually 2 orders of magnitude smaller than $L_\text{k}$.
The parameters $l_\text{ind}$ and $w_\text{ind}$ are fixed design parameters, corresponding to the length and width of the inductor, respectively. 
This procedure allows us to fit $L_g = L_{\text{k},\square}\frac{l_\text{ind}}{w_\text{ind}}$.
To extract $L_{\text{k},\square}$ it is necessary to properly estimate the dimension of the inductor via an SEM of the device.

3. Using the obtained value of $L_{\text{k},\square}$, we calculate the inductance to ground $L_g$ using \eq\eqref{eq:Lk} and then refit the measured spectrum with the correct $L_g$ as a fixed parameter. 
This enables us to determine the values of $C_g$, $\Vec{C_c}$, $C_{i,i+2}/\overline{C}$, and $C_{i,i+3}/\overline{C}$. 
It is important to note that the estimation of parameters using this method is affected by disorder in the CCA, which introduces a small systematic error~\cite{supp}.\\

\paragraph*{\textbf{Disorder estimation: }}
\begin{figure*}
    \centering
    \includegraphics[width = \linewidth]{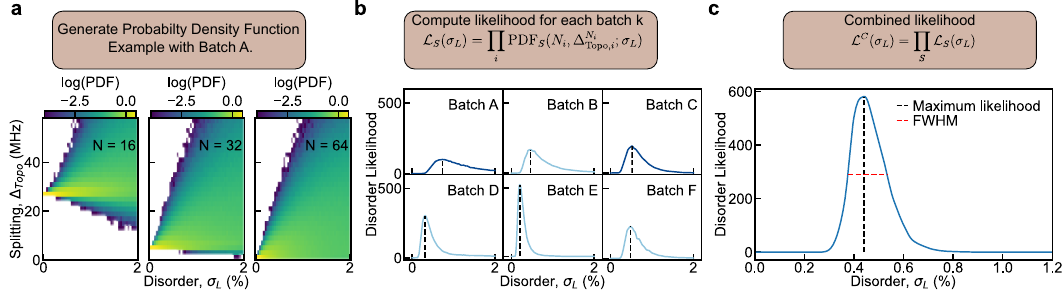}
    \caption{\textbf{Likelihood analysis.} \textbf{a.} Probability density functions of batch $A$ as a function of the disorder, $\sigma_L$, and the splitting between the SSH modes, $\Delta_{\text{Topo}}^N$, for different number of cavities, $N$. \textbf{b.} Likelihood functions computed for each batch using \eq\eqref{methods:eq:likelihood}. 
    The coupling configuration of each batch is represented with the same color code as in \fig\ref{fig:3}.
    The maximum likelihood of each batch is shown with a black dashed lines. 
    \textbf{c.} Combined likelihood between all batches, computed using \eq\eqref{methods:eq:combLikelihood}.
    The maximum likelihood is indicated with a black dashed line, the FWHM is indicated with the red dashed line.
    }
    \label{fig:likelihood}
\end{figure*}
In this section, we outline the procedure for extracting the level of disorder from the frequency splitting of the hybridized SSH modes, $\Delta_\text{Topo}^N$.

The study is performed on six different batches realized in different fabrication runs.
Each batch comprises three to six devices with different number of resonators.
Two batches, A and C, are designed to have a coupling ratio $J_2/J_1 \approx 1.22$, while the others batches are designed to have $J_2/J_1 \approx 1.57$.
We observed changes in $L_{\text{k},\square}$ due to potential fluctuations in the film deposition process (thickness or composition), resulting in up to $15\%$ change of the CCA resonant frequency, $\omega_r/2\pi$.
Those change have a minimal effect on $J_2/J_1$.

We initially employ the fitting routine presented in the previous Methods section to extract the mean capacitances and inductances specific to each batch, assuming a disorder-free scenario.

Then we utilize these parameters in the Hamiltonian \eq\eqref{methods:hamMat} where we introduce Gaussian noise with a standard deviation $\sigma_L$ applied to the inductances:
\begin{equation}
    \bm{H_n}/\hbar = \sqrt{[C^{-1}][L^{-1}\left(\sigma_L\right)]}.
    \label{methods:hamDis}
\end{equation}
Due to the high kinetic inductance of the films and the small size of the inductors, the inductances are sensitive to fabrication imperfection.
Hence, the Gaussian noise, $\sigma_L$, is applied to the inductances of the cavities.
For each batch, number of resonators $N$ and disorder value $\sigma_L$, we generate 30,000 realizations of the Hamiltonian \eq\eqref{methods:hamDis}.
We then diagonalize each of the Hamiltonians and extract $\Delta_\text{Topo}^N$. 
Using these simulations, we construct for each batch a three-dimensional probability density function ($\text{PDF}$) that depends on the number of resonators, the SSH-modes splitting and the level of disorder (\fig\ref{fig:likelihood}\textbf{a}).

Several insights can be derived from these $\text{PDF}$.
First, as expected, the splitting between the SSH modes decreases with an increasing number of resonators.
Secondly, the $\text{PDF}$ exhibit an asymmetry which tends to increase the splitting as a function of disorder.
This asymmetry arises from the fact that disorder can only increase the splitting between the SSH modes.
However, it is noteworthy that for short CCAs, such as the case with 16 resonators shown in \fig\ref{fig:likelihood}\textbf{a}, the splitting can also decrease with increasing disorder. 
This occurs when the SSH modes enter the bulk for sufficiently large disorder values.
Thirdly, when $\Delta_{\text{Topo}}^N$ approaches zero, deviation from its expected value become significantly more prominent.

We proceed to compute the likelihood for each batch $S$, using an interpolated $\text{PDF}$, defined as follows:
\begin{equation}
    \mathcal{L}_{S}(\sigma_L) = \prod_{i}\text{PDF}_S(N_i,\Delta_{\text{Topo},i}^{N_i};\sigma_L)
    \label{methods:eq:likelihood}
\end{equation}
where $N_i$ and $\Delta_{\text{Topo},i}^{N_i}$ represent the number of resonators and the SSH modes frequency splittings of data point (device) $i$, respectively. 
$\text{PDF}_S$ is the Probability density function used for batch $S$.
The likelihood functions (Fig.~\ref{fig:likelihood}\textbf{b}) are then normalized by their area from which we extract the full width at half maximum of the different likelihoods.

To obtain an overall assessment of disorder across all devices of different batches, we employ the method of combined likelihood, defined as follows:
\begin{equation}
    \mathcal{L}^C(\sigma_L) = \prod_S\mathcal{L}_S(\sigma_L),
    \label{methods:eq:combLikelihood}
\end{equation}
where $S$ represents the label of the batch. This function characterizes the typical disorder among all fabricated devices in the topological configuration and is illustrated in Fig.~\ref{fig:likelihood}\textbf{c}.
It is important to note that this method also presents some limitations, as it is sensitive to the accuracy of the estimation of the CCA parameters. 
Errors in the estimation of the coupling capacitances, for example, can lead to significant changes in the decay of the SSH modes. 
One way to mitigate this sensitivity is to operate in a regime where the hybridization between the modes is weak, reducing the impact of parameter misestimation.

\bibliographystyle{apsrev4-2}
\bibliography{Metamaterial}

\appendix

\section{Model of CCAs with $M=1$ \label{app:vanMod}}

In this section, we perform approximations on the derivation of the Hamiltonian presented in Methods \sect\ref{sec:methods}, which allows the implementation of an initial simplified tight-binding interpretation (only including first nearest neighbor interaction) of the CCAs. 
We then enrich the model to include higher-neighbor coupling terms and study their effect on the band structure.

From the model developed in Methods, we start with a CCA with $N$ resonators with $M=1$, assuming a CCA with constant coupling and resonant frequency.
We define the capacitance to ground as $C_g$, the inductance to ground as $L_g$, and the mutual capacitances between resonators as $C_1$.
We proceed to two assumptions on the capacitance matrix developed in Methods (\eq\eqref{methods:eq:numCapa}). 
The first one is to neglect the stray capacitance between resonators higher than their first neighbors, $C_{i,i+j} = 0$ for $j > 1$.
The second assumption has to do with the inversion of the capacitance matrix, where the terms $C_{i,i+j}C_{i+j,i}$ are neglected for orders higher than 1, valid when $C_{i,i+1}/C_{\Sigma,i}$ is small as mentioned in the main text.
Doing so, the inverse of the capacitance matrix becomes
\begin{equation}
    \left[ C^{-1} \right] = 
    L_g\omega_r^2
    \begin{pmatrix}
        1 & \beta & 0 & \hdots & \hdots & 0\\
        \beta & 1 & \beta & \ddots &\ddots& \vdots\\
        0 & \beta & 1 & \beta & \ddots & \vdots\\
        \vdots & \ddots & \ddots & \ddots & \ddots & \vdots\\
        \vdots & \ddots & \ddots & \ddots & \ddots & \beta\\
        0 & \hdots & \ddots & \ddots & \beta & 1
    \end{pmatrix},
    \label{app:eq:vanCapa}
\end{equation}
where we have
\begin{align}
    \beta &= \frac{C_1}{C_\Sigma},\\
    \omega_r &= \frac{1}{\sqrt{L_gC_\Sigma}}.
\end{align}
$C_\Sigma$ is the total capacitance of the resonator defined as $C_\Sigma = C_g + 2 C_1$.
With this simplified capacitance matrix, the Hamiltonian can now be written using \eq\eqref{methods:eq:HamLC} of Methods as
\begin{equation}
    \begin{split}
        H = &\frac{1}{2}\sum_{n = 1}^{N}\left(L_g\omega_r^2Q_n + \frac{1}{L_g}\phi_n \right)\\
        &+\frac{L_g\omega_r^2\beta}{2}\sum_{n=1}^{N-1}\left( Q_nQ_{n+1}+Q_{n+1}Q_n \right).
        \label{app:eq:classicHamiltonian}
    \end{split}
\end{equation}
We now quantize this Hamiltonian, assuming the commutation relation $\left[ \hat{Q}_n, \hat{\phi}_m \right] = i\hbar\delta_{nm}$~\cite{vool2017introduction} which allow to define,
\begin{align}
    \hat{Q}_n &= \sqrt{\frac{\hbar}{2L_g\omega_r}}\left( \hat{a}_n^\dagger + \hat{a}_n \right) \label{app:eq:Q_quant},\\
    \hat{\phi}_n &= \sqrt{\frac{\hbar L_g\omega_r}{2}}\left( \hat{a}_n^\dagger - \hat{a}_n \right)\label{app:eq:Phi_quant},
\end{align}
where $\hat{a}_n^\dagger$ ($\hat{a}_n$), is the annihilation (creation) operators at site $n$.
Including the above relations \eq\eqref{app:eq:Q_quant} and \eq\eqref{app:eq:Phi_quant} in the Hamiltonian, \eq\eqref{app:eq:classicHamiltonian}, we get
\begin{equation}
    \begin{split}
        \hat{H} = &\sum_{n = 1}^{N}\hbar\omega_r\left(\hat{a}_n^\dagger\hat{a}_n + \frac{1}{2} \right)\\
        &+ \hbar J \sum_{n = 1}^{N-1} \left( \hat{a}_n^\dagger\hat{a}_{n+1} + h.c. \right),
    \end{split}
    \label{app:eq:vanHam}
\end{equation}
where 
\begin{equation}
    J = \frac{\omega_r}{2}\frac{C_1}{C_\Sigma},
    \label{app:eq:vanCoupling}
\end{equation} 
is the coupling between the cavities.
In this case, the tight-binding model predicts the emergence of a passband centered around the bare resonance frequency of a single cavity ($\omega_r = 1/\sqrt{L_g (C_g + 2 C_1)}$) with a span of $4J$ with $N$ modes.

\begin{figure}
    \centering
    \includegraphics[width = \linewidth]{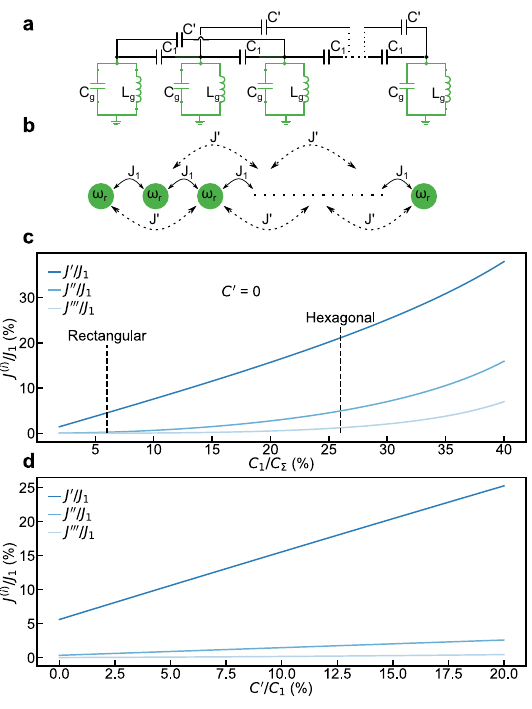}
    \caption{\textbf{Simulation of higher order coupling terms.} \textbf{a.} Circuit schematic of the CCA under consideration. The resonators, highlighted in green, with capacitance $C_g$ and inductance $L_g$ to ground, respectively, are coupled to the first nearest neighbor (NN) with the capacitances $C_1$. We represent the second NN coupling capacitance with $C^\prime$. \textbf{b.} Equivalent CCA schematic. The resonators are represented as cavities with resonant frequency $\omega_r/2\pi$. The first and second NN coupling are respectively indicated with $J_1$ and $J^\prime$. \textbf{c.} Coupling ratio $J^{(i)}/J_1$ of the coupling to the $i^\text{th}$ NN over $C_1/C_\Sigma$. The evolution of second NN coupling $J^\prime$, third NN coupling $J^{\prime\prime}$ and fourth NN coupling $J^{\prime\prime\prime}$ are reported. The two dashed lines indicate the typical $C_1/C_\Sigma$ for the rectangular (\fig\ref{fig:1}\textbf{b}) and hexagonal designs (\fig\ref{fig:1}\textbf{d}). \textbf{d.} $J^{(i)}/J_1$ as a function  of $C^\prime/C_1$. The simulations in panel \textbf{c} and \textbf{d} are performed using \eq\eqref{methods:hamMat} in Methods \sect\ref{sec:methods}.} 
    \label{app:fig:higherOrderCoupling}
\end{figure}

Now, we investigate how coupling terms above first neighbor modify this simplified model. 
Higher order coupling terms will generate terms of type
\begin{equation}
    \hat{K}_{q} = \sum_{n = 1}^{N-q} \hbar J^{(q)} \left( \hat{a}_n^\dagger\hat{a}_{n+q} + h.c. \right),
\end{equation}
where $q$ is the order of the coupling term. 
$J^{(q)}$ represents coupling terms to the $q^{\text{th}}$ nearest neighbor cavity.
In the following, we refer to terms of order $q = 2$, $3$ and $4$ as $J^\prime$, $J^{\prime\prime}$ and $J^{\prime\prime\prime}$, respectively.
These higher coupling terms will simply add up to the real space Hamiltonian \eq\eqref{app:eq:vanHam} as
\begin{equation}
    \hat{H}_\text{tot} = \hat{H} + \sum_{i=2}^I \hat{K}_i,
    \label{app:eq:vanHam_2NN}
\end{equation}
up to the $I^\text{th}$ order.
These higher coupling terms can arise due to the two origin mentioned above, meaning higher order terms due to the inversion of the capacitance matrix and direct stray mutual capacitance.
In \fig\ref{app:fig:higherOrderCoupling}, we show the scaling of these effects on the next nearest neighbor coupling using the Hamiltonian introduced in the Methods section. 

In \fig\ref{app:fig:higherOrderCoupling}\textbf{c}, we show the effect of the increase of the relative coupling capacitance with respect to the total capacitance, $C_1/C_\Sigma$, on the relative higher $m^{\text{th}}$ order photonic hopping rates, $J^{(m)}/J_1$, when the stray mutual capacitance, $C^\prime = C_{i,i+2}$, is set to $0$.
As the relative coupling capacitance is increased, higher-order terms become more important.
In the figure, we highlight the typical coupling ratios, $C_1/C_{\Sigma}$ extracted for the rectangular (6\%) and hexagonal (26 \%) CCA. 
The higher neighbor coupling terms in the hexagonal CCA mainly arise due to a high ratio $C_1/C_\Sigma$.
In \fig\ref{app:fig:higherOrderCoupling}\textbf{d}, we show the effect of the relative stray capacitance, $C^\prime/C_1$, on $J^{(m)}/J_1$. 
$J^\prime/J_1$ increases linearly, as expected from \eq\eqref{app:eq:vanCoupling}.
Additionally, we see that as $C^\prime$ increases, higher and lower order coupling terms also start to emerge.

In summary, for the rectangular CCA geometry, the second nearest neighbour couplings $J^\prime\approx 10\% J_1$ (dominated by direct stray capacitive coupling), while for the hexagonal one, $J^\prime\approx22\% J_1$ (due to high $C_1/C_\Sigma$ ratio).
It is worth remarking that a tight-binding Hamiltonian, which includes only the first two terms in \eq\eqref{app:eq:vanHam_2NN}, represents a valid approximation only when the ratio $J^\prime/J_1$ is negligible.

\begin{figure}
    \centering
     \includegraphics[width =\linewidth]{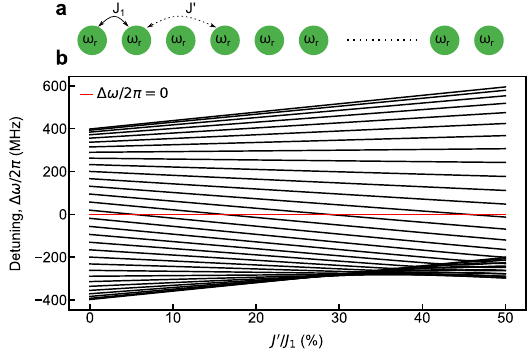}
     \caption{\textbf{Simulation of the influence of second nearest neighbor coupling.}  \textbf{a.} Schematic of a uniform CCA with cavity resonant frequency $\omega_r/2\pi$, coupling to first nearest neighbor (NN) $J_1$ and coupling to second NN $J^\prime$. \textbf{b.} Eigenvalues for a CCA with $N=32$ and couplings typical of a rectangular geometry, calculated according to \eq\eqref{app:eq:vanHam_2NN} as a function of $J^\prime/J_1$. The modes frequencies are represented as a function of the detuning $\Delta\omega = \omega-\omega_r$. The red line represents $\omega = \omega_r$.}
     \label{app:fig:2NN}
\end{figure}

Applying periodic boundary conditions to \eq\eqref{app:eq:vanHam_2NN}, we can write this Hamiltonian in momentum space ($k$-space), using the Fourier transform of the real space creation operator at site $n$,
\begin{equation}
    \hat{a}_n = \frac{1}{\sqrt{N}}\sum_{k}e^{jknd}\hat{a}_k,
    \label{app:eq:a_k}
\end{equation}
where $d$ is the size of the unit cell and $k$ is the wavevector. 
From the equation above we find the $k$-space Hamiltonian,
\begin{equation}
\begin{split}
    H(k) &= \sum_k \hbar\left(\omega_r + 2J^\prime \cos\left(2kd\right) \right) \hat{a}_k^\dagger\hat{a}_k\\ &+ 2\hbar J \hat{a}_k^\dagger\hat{a}_{k+1}\cos\left(kd\right).
    \label{app:eq:dimerHamK}
\end{split}
\end{equation}
The corresponding dispersion relation reads,
\begin{equation}
    \omega_k = \omega_r + 2J_1\cos\left(kd\right) + 2J^\prime\cos\left(2kd\right).
    \label{app:eq:dimerOmega}
\end{equation}
In the limit where $J^\prime = 0$, we find the standard cosine dispersion relation of the tight-binding model, spanned over $4J$ and centered around $\omega_r$.
However, when $J^\prime \neq 0$, significant deviations between the tight-binding model spectrum and this model become evident. 
It manifests in an asymmetric passband with respect to $\omega_r$, resulting in higher mode density at lower frequencies and the eigenmodes of the CCA shift upward with respect to $\omega_r$(\fig\ref{app:fig:2NN}).
The effect on the dissipations in the passband is studied in~\app\ref{app:Dissipations}.

\section{Gapped CCAs with $M \geq 2$ \label{app:bandgap}}
In this section, we model CCAs with gapped spectrum and study, as in the previous section, how they are affected by higher neighbor coupling terms.
In order to open bandgaps, several methods are possible relying on the generation of unit cells~\cite{mirhosseiniSuperconductingMetamaterialsWaveguide2018,liuQuantumElectrodynamicsPhotonic2017}.
Depending on the coupling configuration inside the unit cell, the number of cavities per unit cell will define the number of bandgaps: for $M$ cavities per unit cell, there will be $M-1$ bandgaps.

We first show the case for a dimerized CCA ($M = 2$), which is defined by alternating the mutual capacitances $C_1$ and $C_2$, while keeping the same inductance to ground, $L_g$, and constant resonance frequency among the resonators(see \sect\ref{sec:bandEngineering} of the main text).
In this case, the inverse of the capacitance matrix (\eq\eqref{methods:eq:numCapa}) can be rewritten as,
\begin{equation}
    \left[ C^{-1} \right] = 
    L_g\omega_r^2
    \begin{pmatrix}
        1 & \beta_1 & 0 & \hdots & \hdots & 0\\
        \beta_1 & 1 & \beta_2 & \ddots &\ddots& \vdots\\
        0 & \beta_2 & 1 & \beta_1 & \ddots & \vdots\\
        \vdots & \ddots & \ddots & \ddots & \ddots & \vdots\\
        \vdots & \ddots & \ddots & \ddots & \ddots & \beta_1\\
        0 & \hdots & \ddots & \ddots & \beta_1 & 1
    \end{pmatrix},
\end{equation}
where the stray next nearest neighbor coupling capacitances and the higher order terms in $C_1C_2,C_1^2, C_2^2$ have been neglected, as for \eq\eqref{app:eq:vanCapa} in \app\ref{app:vanMod}, and
\begin{equation}
    \beta_i = C_i/C_\Sigma.
\end{equation}
We can then rewrite the Hamiltonian using \eq\eqref{methods:eq:HamLC} as
\begin{equation}
    \begin{split}
        H = &\frac{1}{2}\sum_{n = 1}^{N}L_g\omega_r^2\left[\left(Q_n^A\right)^2 + \left(Q_n^B\right)^2\right]\\
        &+ \frac{1}{2}\sum_{n = 1}^{N}\frac{1}{L_g}\left[\left(\phi_n^A\right)^2 + \left(\phi_n^B\right)^2\right]\\
        &+\frac{L_g\omega_r^2\beta_1}{2}\sum_{n=1}^{N-1}\left( Q_n^AQ_{n}^B+Q_{n}^AQ_n^B \right)\\
        &+\frac{L_g\omega_r^2\beta_2}{2}\sum_{n=1}^{N-1}\left( Q_n^AQ_{n+1}^B+Q_{n+1}^AQ_n^B \right).
        \label{app:gap:eq:HamSG}
    \end{split}
\end{equation}
We quantize this Hamiltonian by introducing the quantized charge, $\hat{Q}_n^{S}$, and flux, $\hat{\phi}_n^{S}$, operators, acting on the $S^{\text{th}}$ sub-lattice site of the $n^{\text{th}}$ unit cell, satisfying the commutation relation,
\begin{equation}
    \left[\hat{Q}_n^{S^\prime},\hat{\phi}_m^{S}\right] = i\hbar\delta_{n,m}\delta_{S,S^\prime}.
\end{equation}
They are defined as,
\begin{align}
    \hat{Q}_n^S &=\sqrt{\frac{\hbar}{2L_g\omega_r}}\left(\hat{s}_n^\dagger + \hat{s}_n\right) \label{app:gap:eq:Qq}\\
    \hat{\phi}_n^S &=\sqrt{\frac{\hbar L_g\omega_r}{2}}\left(\hat{s}_n^\dagger - \hat{s}_n\right) \label{app:gap:eq:Phiq},
\end{align}
where $\hat{s}_n^\dagger$ ($\hat{s}_n$) is the annihilation (creation) operator of sub-lattice $S$ of the $n^{\text{th}}$ unit cell.\\
Inserting \eq\eqref{app:gap:eq:Qq} and \eq\eqref{app:gap:eq:Phiq} into \eq\eqref{app:gap:eq:HamSG}, we find the dimerized Hamiltonian,
\begin{equation}
    \begin{split}
        \hat{H} &= \hbar\omega_r\sum_{n=1}^N\left(\hat{a}_n^\dagger\hat{a}_n + \hat{b}_n^\dagger\hat{b}_n\right)\\
        &+\underbrace{\hbar J_1\sum_{n=1}^N\left(\hat{a}_n^\dagger\hat{b}_n + \hat{a}_n\hat{b}_n^\dagger\right)}_{\text{Intracell coupling}}\\
        &+ \underbrace{\hbar J_2 \sum_{n=1}^{N-1}\left(\hat{a}_{n+1}^\dagger\hat{b}_n + \hat{a}_{n+1}\hat{b}_n\right)}_{\text{Intercell coupling}}.
    \end{split}
    \label{app:eq:dimerHamiltonian}
\end{equation}
A higher neighbor coupling term can be added similarly to the approach in \app\ref{app:vanMod}. 

\begin{figure*}
    \centering
     \includegraphics[width =\linewidth]{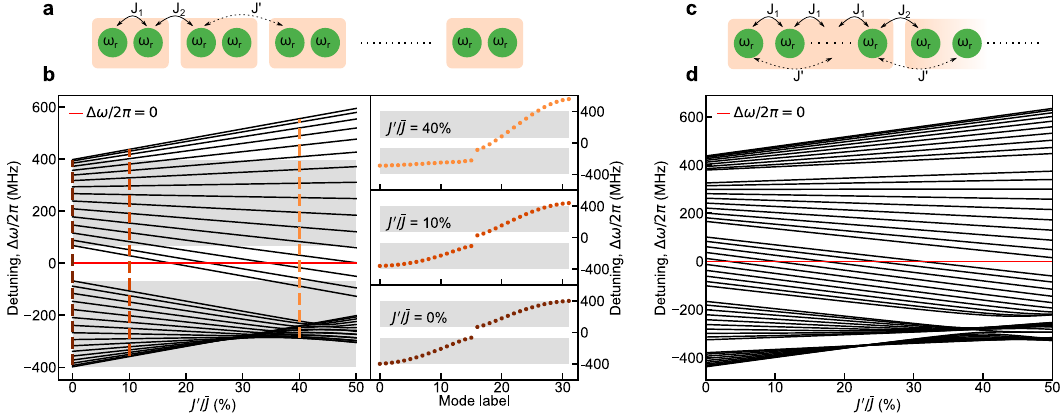}
     \caption{\textbf{Simulation of the influence of second neighbor coupling on gapped CCAs.} \textbf{a.} Schematic of a dimerized ($M=2$) CCA with cavity resonant frequency $\omega_r/2\pi$, intracell coupling $J_1$, intercell coupling $J_2$ and coupling to second nearest neighbor (NN) $J^\prime$. \textbf{b.} Left: Eigenvalues for a CCA with $N=32$ and couplings typical of a rectangular geometry, calculated according to \eq\eqref{app:eq:dimerHamiltonian} as a function of $J^\prime/\overline{J}$, $\overline{J} = 1/2(J_1+J_2)$. The modes frequencies are represented as a function of the detuning $\Delta\omega = \omega-\omega_r$. The red line represents $\omega = \omega_r$. The grey areas highlight the passbands at $J^\prime/\overline{J} = 0$. Right: Cut of the plot in the left panel with $J^\prime/\overline{J} = 0 \%$, $10\%$ and $40\%$. \textbf{c.} Schematic of a multigap ($M>2$) CCA with cavity resonant frequency $\omega_r/2\pi$, intracell couplings $J_1$, intercell coupling $J_2$ and coupling to second nearest neighbor (NN) $J^\prime$. \textbf{d.} Eigenvalues for a CCA with $N=50$ and couplings typical of a rectangular geometry, calculated according to \eq\eqref{app:eq:HamMmer} as a function of $J^\prime/\overline{J}$, $\overline{J} = 1/5(4J_1 + J_2)$. The modes frequencies are represented as a function of the detuning $\Delta\omega = \omega-\omega_r$. The red line represents $\omega = \omega_r$.}
     \label{app:fig:2NN_gap}
\end{figure*}

Analogously to the procedure described in \app\ref{app:vanMod}, one can derive the k-space Hamiltonian to get a grasp of the effect of second neighbor coupling to the mode structure.
To do so, we use the Fourier transform of the normal space field operator, for both $A$ and $B$ sublattice, as in \eq\eqref{app:eq:a_k}.
We find the following Hamiltonian in reciprocal space,
\begin{equation}
\begin{split}
    \hat{H}(k) = \hbar\sum_k\Bigg[\left(\omega_0+2J^\prime\cos{\left(kd\right)}\right)\left(\hat{a}_k^\dagger\hat{a}_k + \hat{b}_k^\dagger\hat{b}_k\right)\\
    \left(J_1+J_2e^{-ikd}\right)\hat{a}_k^\dagger\hat{b}_k + \left(J_1+J_2e^{ikd}\right)\hat{a}_k\hat{b}_k^\dagger\Bigg].
\end{split}
\label{app:eq:hamDimer_kspace}
\end{equation}
By diagonalizing the above Hamiltonian, we recover the dispersion relation,
\begin{equation}
    \omega_\pm/\hbar = \omega_0 + 2J^\prime\cos{\left(kd\right)}\pm\sqrt{J_1^2+J_2^2+2J_1J_2\cos{\left(kd\right)}}.
\end{equation}
Here again, we can see that $J^\prime$ will have a $k$-dependent effect on the band dispersion. 
We simulate its effect on the mode distribution in \fig\ref{app:fig:2NN_gap}\textbf{b}. 
Increasing second neighbor coupling compresses the lower band and dilates the upper band.
This effect can create strong band asymmetry and could be a way to engineer extremely high-density bands. 

One can generalize the model previously introduced for multiple gap systems ($M>1$), where we follow the same recipe as before:
\begin{equation}
    \begin{split}
        \hat{H} &= \hbar\omega_r\sum_{n=1}^{N/M}\sum_m^M\hat{a}_{m,n}^\dagger\hat{a}_{m,n}\\
        +&\hbar \sum_{n=1}^{N/M}\sum_{m=1}^{M-1}J_{m,m+1}\left(\hat{a}_{m,n}^\dagger\hat{a}_{m+1,n} + h.c.\right)\\
        +&\hbar \sum_{n=1}^{N/M-1}J_{M,1}\left(\hat{a}_{M,n}^\dagger\hat{a}_{1,n+1} + h.c.\right),
    \end{split}
    \label{app:eq:HamMmer}
\end{equation}
where $\hat{a}_{m,n}$ is the creation operator on the $m^{\text{th}}$ cavity in the $n^{\text{th}}$ unit cell.

Increasing second neighbor coupling will have the same effect as for the case of single gap devices: compression (expansion) of the lower (upper) bandgaps (\fig\ref{app:fig:2NN_gap}\textbf{d}).

\begin{figure}
    \centering
    \includegraphics[width=\linewidth]{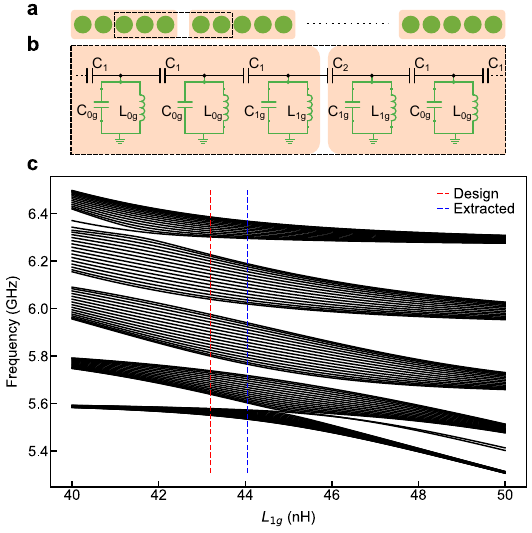}
    \caption{\textbf{Simulation of the influence of systematic inductance deviation in gapped CCA spectra.} \textbf{a.} Schematic of a multigap ($M=5$) CCA. \textbf{b.} Schematic of the lumped-element model of the CCA enclosed by the dashed rectangle in \textbf{a}. The mutual coupling capacitance between the resonators inside the unit cell is $C_1$ and the capacitance between the unit cells is $C_2$. The capacitance and inductance to ground for the resonators inside the bulk of the unit cell are $C_{0g}$ and $L_{0g}$, whereas on the edge of the unit cell they are defined as $C_{1g}$ and $L_{1g}$, respectively. \textbf{c.} Simulation of the modes' frequencies as a function of $L_{1g}$. The red and blue dashed lines highlight the designed and extracted frequency, respectively. The simulation in panel \textbf{c} is performed using the Hamiltonian \eq\eqref{methods:eq:HamLC} in Methods \sect\ref{sec:methods}}
    \label{app:fig:MG-deviation}
\end{figure}

In the measurements presented in \fig\ref{fig:2}\textbf{e} of the main text, we see a clear asymmetry in the bandgap size of the spectra with more than 1 gap. 
This asymmetry can be explained by a systematic deviation of the resonator's inductances of the real device with respect to the designed one.
We attribute this deviation to the resonant frequency difference between the resonators at the edge and in the bulk of the unit cells. 
In the design, this effect was taken into account by having different inductances for edge and bulk resonators of a unit cell (see \app\ref{app:design}). 
This effect is highlighted in \fig\ref{app:fig:MG-deviation}, where we show that errors on the estimation of the inductance as low as $\SI{0.6}{\nano\henry}$, about $1.5\%$ of the total inductance, drastically affects the band structure.

\section{\label{app:design}Design and simulation}

All designs were created in the \textit{.gds} format using the gdspy Python library~\cite{gabrielliGDSPY2023}. 
To ensure the \textit{lumpedness} of the designed resonators, the inductor and capacitor self-resonances' are designed to be much higher than the frequency range we are working in. 
The typical design workflow begins with an initial tight-binding simulation based on the model established in \app\ref{app:vanMod} and \app\ref{app:bandgap}.
Subsequently, we use Sonnet simulation software to simulate the frequency response of the designs. 
In this simulation, the inductance is set by defining the number of squares in the inductor to match the required inductance corresponding to the kinetic inductance of the film. 
From this simulation, we extract the impedance and resonant frequencies of the design.\\
\begin{figure}
    \centering
    \includegraphics[width = \linewidth]{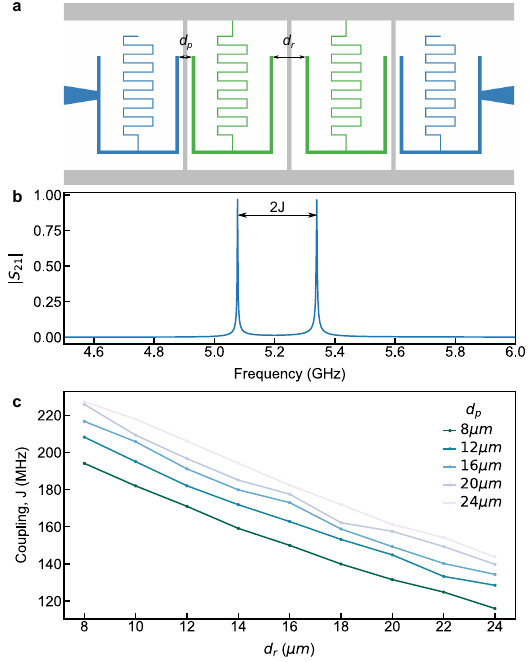}
    \caption{\textbf{Coupling calibration for rectangular CCA geometry.} \textbf{a.} Schematic of the calibration design, with the cavities in green, the coupling ports in blue and the ground plane in grey. $d_p$ is the distance between the ports (ghosts) and cavities, and $d_r$ is the distance between the cavities. \textbf{b.} Simulation using Sonnet software of the transmission amplitude $|S_{21}|$ through the dimer as a function of the excitation frequency. The coupling, $J$, is extracted from the frequency splitting between the two modes frequencies. \textbf{c.} Coupling extracted from Sonnet simulations as a function of $d_r$ for different values of $d_p$, assuming $L_{\text{k},\square} = 100$ pH/$\square$.}
    \label{app:fig:dimerSweep}
\end{figure}
For the rectangular CCAs, the calibration of the coupling rate between resonators is achieved by simulating two resonators positioned adjacent to each other, with a distance denoted as $d_r$ ($d_p$) with respect to the neighboring resonator (microwave feedline), as depicted in \fig\ref{app:fig:dimerSweep}\textbf{a}.

The coupling between two resonators is proportional to $J_{i,i+1} \propto \frac{C_{i,i+1}}{C_\Sigma}$, with $C_{i,i+1}$ being the coupling capacitance and $C_\Sigma$ the total capacitance of the resonator. 
As the distance between the two resonators is increased, the coupling between them is reduced as $C_c$ is decreasing (\fig\ref{app:fig:dimerSweep}\textbf{b}).

\begin{figure}
    \centering
    \includegraphics[width = \linewidth]{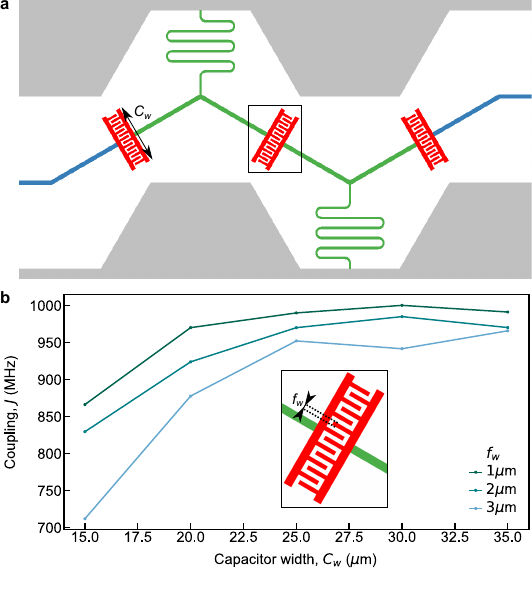}
    \caption{\textbf{Coupling calibration for hexagonal CCA geometry.} \textbf{a.} Schematic of the calibration design, with the cavities in green, the coupling ports in blue, the capacitors that are swept through the simulation in red, and the ground plane in grey. $C_w$ is the coupling capacitor width. \textbf{b.} Coupling extracted from Sonnet simulations as a function of $C_w$ for different values of, $f_w$, representing the finger and gap width. The inset shows a zoom-in of the capacitor from panel \textbf{a}, assuming $L_{\text{k},\square} = 100$ pH/$\square$.}
    \label{app:fig:LHcal}
\end{figure}

For the hexagonal CCAs, the calibration of the coupling rate between resonators is implemented by sweeping the width of the coupling capacitor $C_w$ from 15 to \SI{35}{\micro\metre} for three different finger width equal to the finger spacing $f_w = 1,2,\SI{3}{\micro\metre}$ (see \fig\ref{app:fig:LHcal}).
We then extract the mode splitting from which we infer the coupling between the cavities.

\begin{figure}
    \centering
    \includegraphics[width =\linewidth]{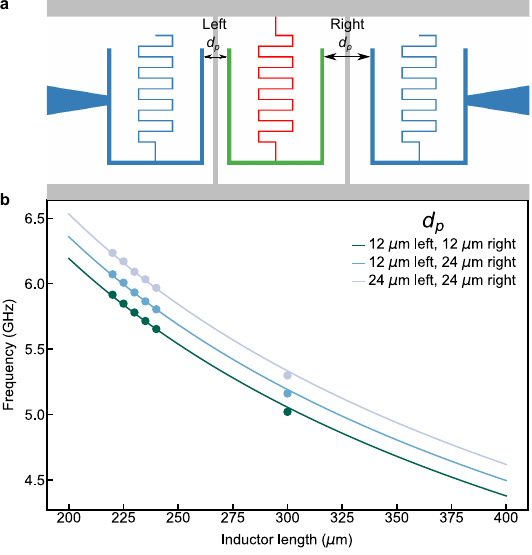}
    \caption{\textbf{Calibration of the resonant frequency for multigap CCAs.} \textbf{a.} Schematic of the calibration design, with the capacitor of the cavity in green, the inductor swept through the simulation in red, the coupling ports in blue, and the ground plane in grey. \textbf{b.} Simulated resonant frequency of the cavity as a function of the length of the inductor for different spacing to coupling ports, $d_p$, using Sonnet software and assuming $L_{\text{k},\square} = 100$ pH/$\square$. The dots represent the resonant frequencies extracted from the simulation and the lines are fits with \eq\eqref{met:eq:fvsLsq}.}
    \label{app:fig:resCalibration}
\end{figure}

For $M=1$ cavities per unit cell, all cavities are coupled to one another with the same mutual capacitance, $C_{i,i+1} = C_1$, have a capacitance to ground, $C_g$, and an inductance to ground, $L_g$.
In this case, all cavities will have the same total capacitance, i.e. $C_\Sigma = Cg + 2C_1$.
This condition ensures that all cavities have the same resonant frequency, $\omega_r = 1/\sqrt{L_gC_\Sigma}$.
Maintaining a constant resonant frequency among all the resonators in the array poses a challenge due to the presence of local fabrication imperfections and edge coupling points to the input-output ports. 
While the former is determined by fabrication capabilities, the latter can be mitigated by introducing "ghost" cavities, as indicated in blue in the SEM micrographs (\figs\ref{fig:1}\textbf{b} and \textbf{d} of the main text). 
Such elements are designed to mimic the capacitive environment of bulk resonators, ensuring uniform resonant frequencies also for the $1^{\rm st}$ and ${\rm N}^{\rm th}$ resonators of the CCA, provided that all resonators share identical inductance values. 
These "ghost" elements are designed to be non-resonant with the CCA and are electrically connected to the input/output waveguide, serving as input/output ports of the CCA with capacitance $C_p\approx C_1$, $C_p$ being the capacitor between the coupling ports and the edge cavities.

For $M=2$, the cavities are coupled in an alternative fashion with the capacitances $C_1$ and $C_2$, and have a capacitance to ground, $C_g$, and an inductance to ground, $L_g$.
Similarly to the case with $M=1$, the cavities will all have the same total capacitance, i.e. $C_\Sigma = C_g + C_1 + C_2$ for each cavity. 
Hence, they will have the same resonant frequency, $\omega_r/2\pi$, provided that the coupling capacitance to the input/output waveguides is equal to $C_2$.

In the dimer case ($M=2$), each resonator presents the same two coupling capacitances, in an alternating fashion, which automatically satisfies the resonant condition. 
However, for $M>2$, the translation uniformity for a single resonator is not preserved anymore. 
Therefore, accomplishing the resonant frequency condition for all cavities requires precise control over the mutual and ground capacitances. 
We address this challenge by utilizing RF simulation software to calibrate the resonant frequency as a function of the coupling strength of each of the cavities in the unit cell.
In fact, for $M>2$, the cavities in the unit cell are coupled to one another with $C_1, C_2, C_3, \dots, C_{M-1}$ and with $C_M$ between unit cells.
They have a capacitance to ground, $C_{1g}, C_{2g}, C_{3g}, \dots, C_{Mg}$ and inductance to ground $L_{1g}, L_{2g}, L_{3g}, \dots, L_{Mg}$.
In this case the cavities within the unit cell will each have a different total capacitance, $C_{\Sigma,i} = C_{ig} + C_{i-1} + C_{i+1}$.
To ensure the same resonant frequency for all cavities, the inductances have to be adjusted such that, $C_{\Sigma,i}L_i = C_{\Sigma,j}L_j$.
In order to simplify the design, the CCAs presented in \fig\ref{fig:2}\textbf{e}, are implemented restricting the mutual capacitances to $C_1$ and $C_2$, the capacitances to ground to $C_{1g}$ and $C_{2g}$ and the inductances to ground to $L_{1g}$ and $L_{2g}$.

For instance, for the CCA with $M=3$ presented in the main text, the unit cell coupling capacitance $\Vec{C_c} = (C_1, C_1, C_2)$.
Hence, here the cavities at the edges of the unit cell will experience a different capacitive environment than the cavities in the bulk of the CCA.
Therefore, we need to adapt the inductances for only two types of cavities: the ones at the edges and the ones in the bulk of the unit cell. 
$\Vec{L_g} = (L_{1g}, L_{2g})$ needs to be adjusted to keep a constant resonant frequency.

\begin{figure*}
    \centering
    \includegraphics[width = \linewidth]{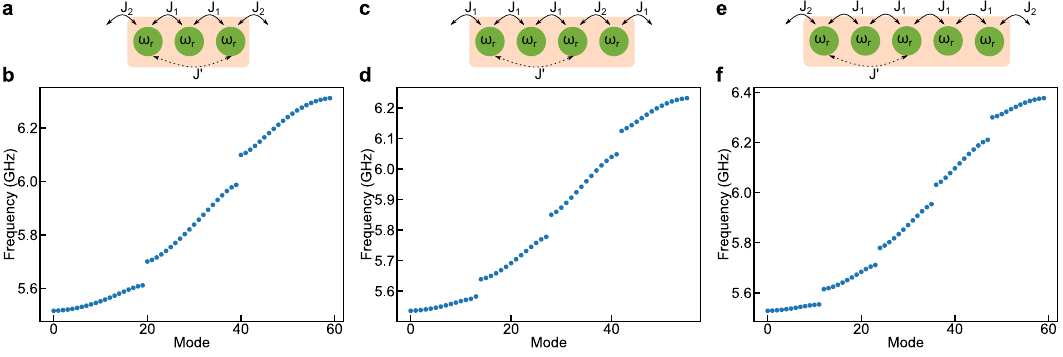}
    \caption{\textbf{Mode dispersion for multigap CCAs.} \textbf{a.} (\textbf{c.}, \textbf{e.}) CCA schematic of the unit cell of multigap designs used in \fig\ref{fig:2}\textbf{e} with $M = 3$ ($4$, $5$) cavities per unit cell, resulting in 2 (3, 4) bandgaps. \textbf{b.} (\textbf{d.}, \textbf{f.}) Simulated mode dispersion for the 2 (3, 4) bandgap devices with $N = 60$ ($56$, $60$) cavities, simulated with the Hamiltonian \eq\eqref{methods:hamMat} presented in Methods \sect\ref{sec:methods}.}
    \label{app:fig:MGdesign}
\end{figure*}

The designs of the CCA with multiple gaps were chosen by testing different coupling configurations.
The simulations were performed using the Hamiltonian introduced in \eq\eqref{app:eq:HamMmer} in \app\ref{app:bandgap}.
The simulated mode profiles of each chosen design measured in \fig\ref{fig:2}\textbf{e} are displayed in \fig\ref{app:fig:MGdesign}.

\section{Dissipations \label{app:Dissipations}}
\subsection{Modelling of the CCA dissipations}
In this section, we focus on the dissipations occurring in the CCA.
We derive a non-Hermitian Hamiltonian and the scattering matrix for the CCA that we implemented.
The presence of non-negligible second neighbor coupling among cavities suggests the possibility of coupling between the input waveguide and the second closest resonator to the microwave ports.
In our subsequent analysis, we incorporate these couplings and observe their effect, noting that they introduce asymmetry in dissipation with respect to frequency detuning.
To model dissipations, we start from the Heisenberg-Langevin equation of motion~\cite{wallsQuantumOptics2008},
\begin{equation}
\begin{split}
    \frac{\partial}{\partial t}\hat{a}_n(t) &= j\left[\hat{H},\hat{a}_n(t)\right] + \frac{\kappa_n}{2}\hat{a}_n(t)\\
    &+ \sqrt{\kappa_\text{ext}}\left(\delta_{n,1}\hat{a}_\text{in,L}(t) + \delta_{n,N}\hat{a}_\text{in,R}(t)\right)\\
    &+ \sqrt{\kappa_\text{ext}^\prime}\left(\delta_{n,2}\hat{a}_\text{in,L}(t) + \delta_{n,N-1}\hat{a}_\text{in,R}(t)\right),
    \label{app:eq:Heisenberg}
\end{split}
\end{equation}
where $\kappa_\text{ext}$ and $\kappa_\text{ext}^\prime$ are the dissipation rates to the coupling ports from the closest and second closest resonator to the microwave ports, respectively. 
$\kappa_n = \kappa_\text{int} + \left(\delta_{n,1} + \delta_{n,N}\right)\kappa_\text{ext} + \left(\delta_{n,2} + \delta_{n,N-1}\right)\kappa_\text{ext}^\prime$ is the total dissipation of the $n^{\text{th}}$ cavity, where $\kappa_\text{int}$ is the cavity's internal dissipation. 
$\hat{a}_\text{in,L(R)}$ is the input field on the left (right) port.
$\hat{H}$ is the Hamiltonian of the system under study; in this case, we work with the normal Hamiltonian \eq\eqref{app:eq:vanHam} for the sake of simplicity.
In the following, we assume that the internal dissipation rate, $\kappa_\text{int}$, is the same for all cavities.

One can write the previous equation in the steady state regime where $\hat{a}_n(t) = \hat{a}_n\exp{\left(-i\omega t\right)}$ and $\hat{a}_\text{in}(t) = \hat{a}_\text{in}\exp{\left(-i\omega t\right)}$, resulting in,
\begin{equation}
\begin{split}
    & \hat{a}_n\left( \Delta - j\frac{\kappa_n}{2}\right) + J\left(\hat{a}_{n+1}\left(1-\delta_{n,N}\right) + \hat{a}_{n-1}\left(1-\delta_{n,1}\right)\right)\\
    & -j\sqrt{\kappa_\text{ext}}\left(\delta_{n,1}\hat{a}_\text{in,L} + \delta_{n,N}\hat{a}_\text{in,R}\right)\\
    & -j\sqrt{\kappa_\text{ext}^\prime}\left(\delta_{n,2}\hat{a}_\text{in,L} + \delta_{n,N-1}\hat{a}_\text{in,R}\right) = 0,
    \label{app:eq:Heisenberg-steady}
\end{split}
\end{equation}
where $\Delta = \omega_r - \omega$ is the detuning between the probe frequency $\omega/2\pi$ and $\omega_r$. We can now use the two input/output relations,
\begin{align}
    \hat{a}_\text{in,L(R)} + \hat{a}_\text{out,L(R)} &= \sqrt{\kappa_\text{ext}}\hat{a}_{1(N)} \label{app:eq:in-out1}\\
    \hat{a}_\text{in,L(R)} + \hat{a}_\text{out,L(R)} &= \sqrt{\kappa_\text{ext}^\prime}\hat{a}_{2(N-1)}\label{app:eq:in-out2}
\end{align}
where $\hat{a}_\text{out,L(R)}$ is the output field on the left (right) side of the CCA. Note that the two input/output relations use the same input/output fields.
By inserting \eqs\eqref{app:eq:in-out1} and \eqref{app:eq:in-out2} into \eq\eqref{app:eq:Heisenberg-steady} one obtains, 
\begin{equation}
\begin{split}
    & \hat{a}_n\left( \Delta - j\frac{\kappa_n}{2}\right) + J\left(\hat{a}_{n+1}\left(1-\delta_{n,N}\right) + \hat{a}_{n-1}\left(1-\delta_{n,1}\right)\right)\\
    &- j\sqrt{\kappa_\text{ext}}\sqrt{\kappa_\text{ext}^\prime}\left(\delta_{n,1}\hat{a}_2 + \delta_{n,2}\hat{a}_1\right)\\
    &- j\sqrt{\kappa_\text{ext}}\sqrt{\kappa_\text{ext}^\prime}\left(\delta_{n,N-1}\hat{a}_N + \delta_{n,N}\hat{a}_{N-1}\right)\\
    &- j\sqrt{\kappa_\text{ext}}\sqrt{\kappa_\text{ext}^\prime}\left(\delta_{n,1}\hat{a}_\text{out,L} + \delta_{n,N}\hat{a}_\text{out,R}\right)\\
    &- j\sqrt{\kappa_\text{ext}}\sqrt{\kappa_\text{ext}^\prime}\left(\delta_{n,2}\hat{a}_\text{out,L} + \delta_{n,N-1}\hat{a}_\text{out,R}\right)= 0,
    \label{app:eq:Heisenberg-steady2}
\end{split}
\end{equation}
This allows us to write down the non-Hermitian Hamiltonian in the field basis $A = \{\hat{a}_1,\hat{a}_2,\hdots,\hat{a}_n\}$,
\begin{widetext}
    \begin{equation}
    \frac{\bm{H_n^\text{Non-Herm}}}{\hbar} = \begin{pmatrix}
        \omega_r - j\frac{\kappa_1}{2}  & J - j\sqrt{\kappa_\text{ext}^\prime}\sqrt{\kappa_\text{ext}} & 0 & \hdots & \hdots & 0 \\
        J - j\sqrt{\kappa_\text{ext}^\prime}\sqrt{\kappa_\text{ext}} & \omega_r - j\frac{\kappa_2}{2} & J & \ddots & \ddots & \vdots \\
        0 & J & \omega_r - j\frac{\kappa_3}{2} & J & \ddots & \vdots\\
        \vdots &\ddots & \ddots & \ddots & \ddots & \vdots \\
        \vdots & \ddots & \ddots & \ddots & \omega_{r} - j\frac{\kappa_{N-1}}{2} & J - j\sqrt{\kappa_\text{ext}^\prime}\sqrt{\kappa_\text{ext}}  \\
        0 & \hdots & \hdots & \hdots & J - j\sqrt{\kappa_\text{ext}^\prime}\sqrt{\kappa_\text{ext}}  & \omega_r - j\frac{\kappa_{N}}{2}
    \end{pmatrix}.
    \label{app:eq:nonHermtianHam}
    \end{equation}
\end{widetext}
Despite this non-Hermitian has been derived for a uniform CCA, this procedure is valid for any CCAs discussed in the manuscript.
We use this non-Hermitian Hamiltonian to extract the dissipation properties of the resonators in the CCA. 
Before moving on to this part, we show below how to obtain the scattering coefficients of any CCA.
One can conveniently define the scattering coefficients as~\cite{naamanSynthesisParametricallyCoupled2022},
\begin{equation}
    S_{kl} = \frac{a_l^\text{out}}{a_k^\text{in}} = j\sqrt{\kappa_{\text{ext,k}}\kappa_{\text{ext,l}}}\left[\bm{M}^{-1}\right]_{kl} - \delta_{kl},
    \label{app:eq:inputOutput}
\end{equation}
where $\kappa_{\text{ext,j}}$ and $\kappa_{\text{ext,l}}$ are the dissipation rates to the output, $k$, and input, $l$, measurement ports, respectively. $\left[\bm{M}\right]$ is the equation of motion matrix defined as,
\begin{equation}
    [\bm{M}] = [\bm{H_n^\text{Non-Herm}}] - \omega\mathds{1}.
\end{equation}
Given our choice of coupling ports, we only measure the scattering parameters $S_{N1}$, $S_{11}$, $S_{NN}$ and $S_{1N}$, which we call with the standard notation, $S_{21}$, $S_{11}$, $S_{22}$ and $S_{12}$.

\subsection{Extracting dissipations}
\begin{figure*}
    \centering
    \includegraphics[width = \linewidth]{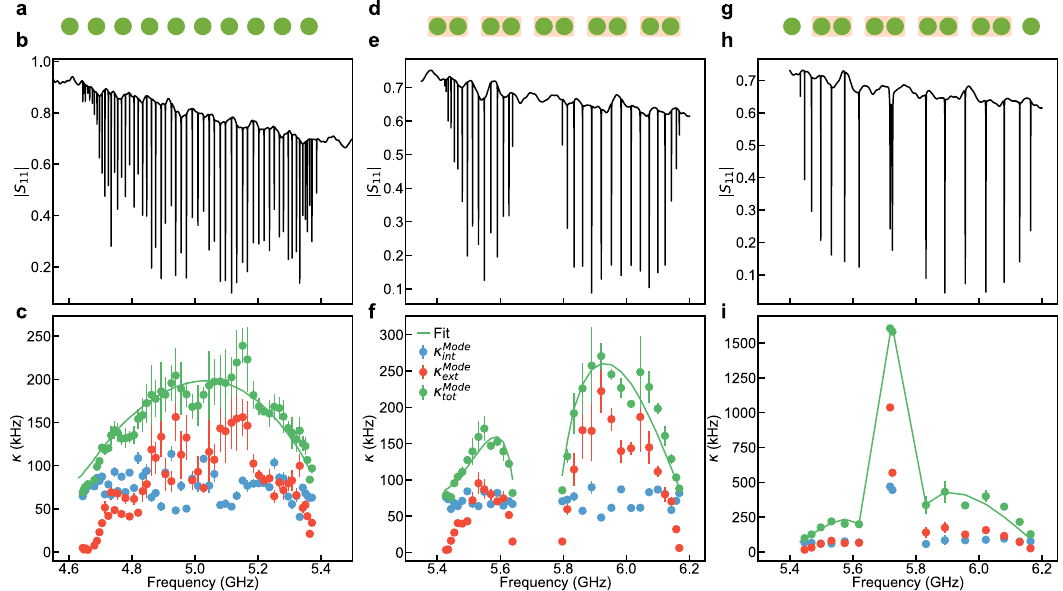}
    \caption{\textbf{Fit of the dissipation for representative CCAs in the normal, trivial and topological configuration.} \textbf{a.} (\textbf{d.}, \textbf{g.}) CCA schematic in the normal (trivial, topological) configuration. \textbf{b.} (\textbf{e.}, \textbf{h.}) Magnitude of the reflection signal, $|S_{11}|$, of a CCA in the normal (trivial, topological) configuration with $N = $ 64 (32,16), measured at low power. \textbf{c.} (\textbf{f.}, \textbf{i.}) Internal (blue), external (red) and total (green) dissipation rates fitted for each mode according to \eq\eqref{app:eq:S11_single}. The green line is a fit of the total dissipation of the modes, $\kappa_\text{tot}^\text{Mode}$, according to the complex part of the eigenvalues of the non-Hermitian Hamiltonian \eq\eqref{app:eq:nonHermtianHam}.}
    \label{app:fig:Dissipation}
\end{figure*}
In order to extract the dissipation parameters of the system, we use both the non-Hermitian Hamiltonian~\eq\eqref{app:eq:nonHermtianHam} and the scattering matrix~\eq\eqref{app:eq:inputOutput}. 
Fitting the full scattering matrix is challenging, due to the number of fitting parameters and the long computational time of the full scattering matrix.
We proceed instead by extracting the dissipations mode by mode, modeling each eigenmode as a single resonator with a certain external coupling ($\kappa_{\text{ext}}^\text{Mode}$) and internal loss rate ($\kappa_{\text{int}}^\text{Mode}$).
The reflection scattering parameter of a single mode can be defined as
\begin{equation}
    S_{11} = 1-\frac{\kappa_{\text{ext}}^\text{Mode}}{i\Delta + (\frac{\kappa_{\text{ext}}^\text{Mode}}{2} + \frac{\kappa_{\text{int}}^{\text{Mode}\star}}{2})} = S_{22},
    \label{app:eq:S11_single}
\end{equation}
where $\kappa_{\text{int}}^{\text{Mode}\star}$ is the extracted internal dissipation from the fit, which also takes into account the dissipation to the other coupled microwave waveguide, at the other end of the CCA.
Hence, we have $\kappa_\text{int} = \kappa_{\text{int}}^{\text{Mode}\star} - \kappa_{\text{ext}}^\text{Mode}$, where $\kappa_\text{int}$ is the internal dissipation to the environments.

We can fit the total modes' dissipation, $\kappa_\text{tot}^{\text{Mode}} = \kappa_\text{int}^{\text{Mode}} + \kappa_\text{ext}^{\text{Mode}}$ with the complex part of the eigenvalues of the non-Hermitian Hamiltonian, \eq\eqref{app:eq:nonHermtianHam}, from which we can extract the internal and external dissipations of the cavities in the CCA.
In \figs\ref{app:fig:Dissipation} (\textbf{c}, \textbf{f} and \textbf{i}), we fit the extracted dissipations for some representative devices in the normal ($J_1 = J_2$), topologically trivial ($J_1 > J_2$) and topologically non-trivial ($J_1 < J_2$) coupling configurations. 
From this fit, we can observe a clear asymmetry in the dissipations to the coupling port: the lower frequency modes (lower passband) are less coupled than the upper frequency modes (upper passband). 
We establish that this effect is caused by, $\kappa_\text{ext}^\prime$ the coupling between the microwave port and the second closest resonator to the microwave ports. 
Even though $\kappa_\text{ext}^\prime \approx \SI{10}{\kilo\hertz}\ll\kappa_\text{ext}$, it has a significant effect on the modes amplitudes.
This effect is also qualitatively observable in all measured transmission spectra in \fig\ref{fig:2}, \fig\ref{fig:3}, \fig\ref{app:fig:deviceNotMeas} and \fig\ref{app:fig:topoFig4}.

\begin{table*}
    \caption{Table of quality factors of representative devices at low photon number. $\overline{\kappa_\text{int}^\text{Mode}}$ and $\overline{\kappa_\text{ext}^\text{Mode}}$ are the mean internal and external dissipation rates of the modes of a CCA. $\kappa_\text{int}$ is the internal dissipation rate of the cavities in the CCA. $\kappa_\text{ext}$ and $\kappa_\text{ext}^\prime$ are the dissipation rates to the coupling ports from the closest and second closest cavity to the microwave ports. $\omega_r$ is the frequency of the cavities in the CCA.}
    \rowcolors{2}{gray!25}{white}
    \begin{tabular}{c|c|c|c|c|c}
     Figure/Device & $\omega_r/\overline{\kappa_\text{int}^\text{Mode}}$ ($\times10^3$) & $\omega_r/\overline{\kappa_\text{ext}^\text{Mode}}$ ($\times10^3$) & $\omega_r/\kappa_\text{int}$ ($\times10^3$) & $\omega_r/\kappa_\text{ext}$ ($\times10^3$)& $\omega_r/\kappa_\text{ext}^{\prime}$ ($\times10^3$)\\
    \hline
    \fig\ref{fig:2}\textbf{c}/Left & 20 & 100 & 11 & 4 & 1242 \\
    \fig\ref{fig:2}\textbf{c}/Middle & 70 & 137 & 10 & 2.32 & 75 \\
    \fig\ref{fig:2}\textbf{c}/Right & 8.6 & 9.5 & 24.6 & 107 & 49 \\
    \fig\ref{fig:2}\textbf{d}/Left & 62 & 63 & 22 & 1.67 & 122 \\
    \fig\ref{fig:2}\textbf{d}/Right & 85 & 120 & 33 & 2.73 & 169 \\
    \fig\ref{fig:2}\textbf{e}/Top & 37 & 116 & 26 & 1.27 & 67 \\
    \fig\ref{fig:2}\textbf{e}/Bottom & 41 & 144 & 26 & 1.5 & 35 \\
    \fig\ref{fig:3}\textbf{d}/16 & 34 & 29 & 12 & 1.48 & 145 \\
    \fig\ref{fig:3}\textbf{f}/16 & 75 & 30 & 35 & 1.1 & 363 
    \end{tabular}
    \label{tab:deviceFig4}
\end{table*}

\begin{figure*}
   \centering
   \includegraphics[width = \linewidth]{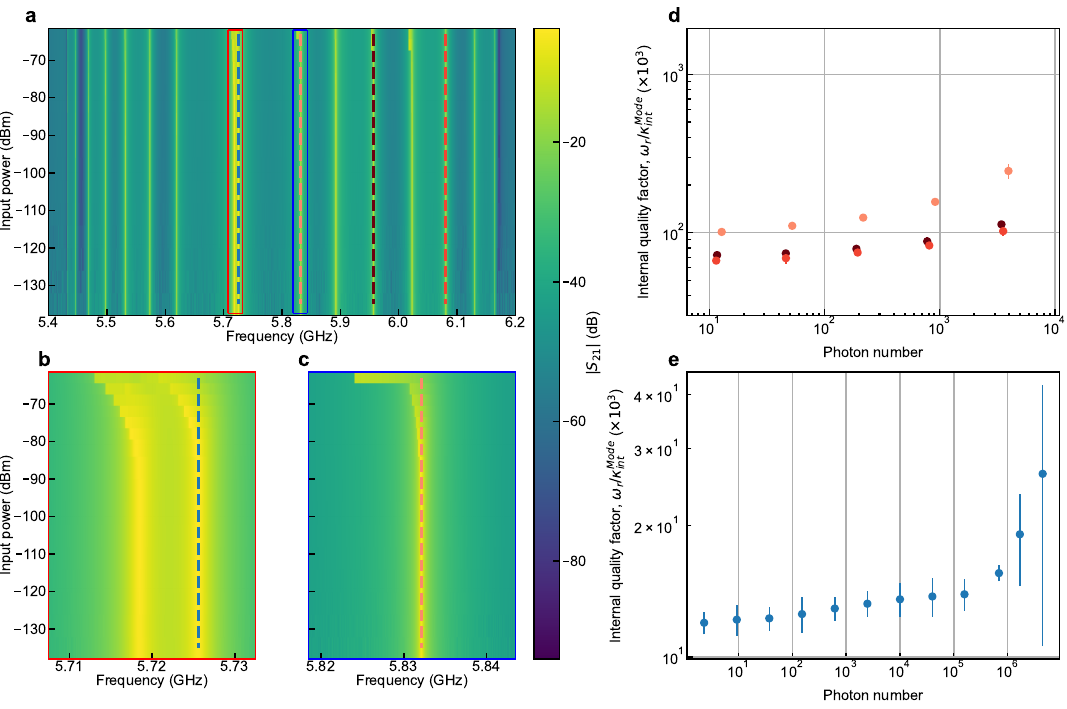}
   \caption{\textbf{Power dependence of the CCAs.} \textbf{a.} Power scan of a CCA in the topological configuration with $J_2/J_1 = 1.57$ and $N = 16$ measured in transmission. \textbf{b.} Zoom-in of the power scan on the SSH modes. \textbf{c.} Zoom-in of the power scan on a mode of the bulk. The frequency span for both cuts is the same. \textbf{d.} Internal quality factors, $\omega_r/\kappa_\text{int}^\text{Mode}$, of representative bulk modes as a function of the estimated photon number in the modes. The fitted modes are highlighted by the three red dashed lines in panels \textbf{a} and \textbf{c}. \textbf{e.} Internal quality factors, $\omega_r/\kappa_\text{int}^\text{Mode}$, of a representative SSH mode as a function of the estimated photon number. The mode is highlighted by the blue dashed lines in panels \textbf{a} and \textbf{b}.}
   \label{app:fig:powScan}
\end{figure*}

\section{SSH \label{app:SSH}}

SSH states represent a unique category of symmetry protected topological states that manifest in 1D systems characterized by alternating hopping amplitudes. 
They have been originally introduced in the realm of condensed matter physics to describe the electronic structure of polyacetylene chains, a 1D organic polymer \cite{suSolitonExcitationsPolyacetylene1980}.
For a careful derivation of the SSH model and discussion of its properties, we refer the reader to~\cite{dalibardCollegeFranceLecture2018,asbothShortCourseTopological2016}.

\subsection{SSH model and influence of second neighbor coupling}

In this section, we numerically model the devices measured in \fig\ref{fig:3}.
Specifically, we study the evolution of the energy spectra of the SSH CCA as a function of the next nearest neighbor coupling, $J^\prime$, and the number of resonators, $N$.
The CCA in the SSH coupling configuration is described by the Hamiltonian \eq\eqref{app:eq:hamDimer_kspace} with $J_1<J_2$.
With the CCA in the SSH configuration and $J^\prime = 0$, we expect the formation of bulk bands separated by $2|J_2-J_1|$ and the presence of in-gap modes at the center of the bandgap physically localized at the edges of the CCA.

\begin{figure}
    \centering
    \includegraphics[width=\linewidth]{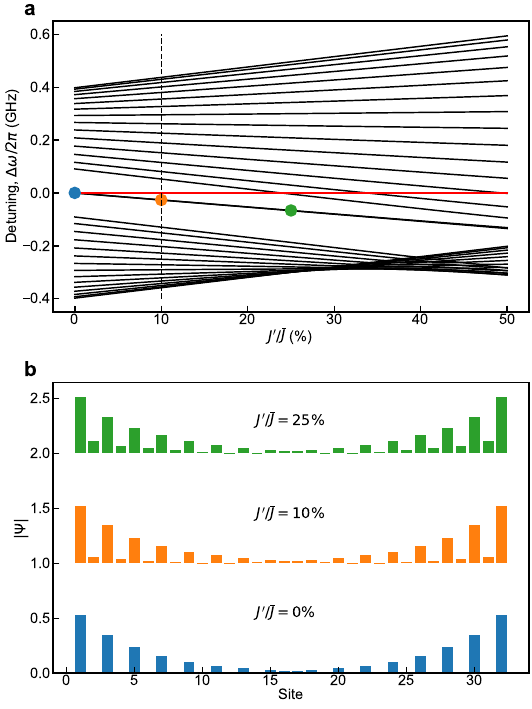}
    \caption{\textbf{Influence of second neighbor coupling on the SSH modes.} \textbf{a.} Eigenvalues for a CCA with $N=32$ and $M=2$ in the topologically non-trivial coupling configuration, calculated according to \eq\eqref{app:eq:dimerHamiltonian} as a function of $J^\prime/\overline{J}$, where $\overline{J} = 1/2(J_1+J_2)$. The modes frequencies are represented as a function of the detuning $\Delta\omega = \omega-\omega_r$. The red line represents $\omega = \omega_r$. For this simulation we have used, $J_1/2\pi = \SI{160}{\mega\hertz}$, $J_2/2\pi = \SI{240}{\mega\hertz}$ and $\omega_r/2\pi = \SI{5}{\giga\hertz}$. The vertical dashed line indicates the typical $J^\prime/\overline{J}$ for the rectangular CCA geometry. \textbf{b.} Simulation results depicting the norm of the spatial distribution of the symmetric SSH mode for various strengths of second neighbor coupling are presented in \textbf{a}.}
    \label{app:fig:SSH2NN}
\end{figure}

We are now going to consider the $J^\prime\neq0$ term in \eq\eqref{app:eq:hamDimer_kspace} and study its effects on the spectrum of the CCAs.
In \eq\eqref{app:eq:hamDimer_kspace} of the main text, we observe that the action of $J^\prime$ in the Hamiltonian is proportional to $\tau_0$, for $J^\prime/\overline{J} < 30\%$ it should not influence qualitatively the topological properties of the system~\cite{perez-gonzalezInterplayLongrangeHopping2019}.
We first study how $J^\prime\neq0$ modifies the SSH CCA spectrum (\fig\ref{app:fig:SSH2NN}).
As already observed in the topologically trvial case (see \fig\ref{app:fig:2NN_gap}) we observe a compression of the low passband together with an expension of the upper passband as a function of $J^\prime$. 
In addition, while increasing $J^\prime$ the SSH modes are also shifted with respect to the resonant frequency of the CCA resonators, $\omega_r/2\pi$ (see red line in \fig\ref{app:fig:SSH2NN}), up to the point where they cannot distinguished from the bulk modes of the upper passband.
$J^\prime$ also impacts the modes' spatial distribution, allowing for some photonic population to extends into the neighboring sub-cell, thereby breaking chiral symmetry, even in the absence of a $\tau_z$ term in the Hamiltonian in \eq\eqref{app:eq:hamDimer_kspace}. 
This effect becomes stronger as $J^\prime$ is increased (\fig\ref{app:fig:SSH2NN}).

\subsection{Amplitude of the SSH modes}
\begin{figure}
    \centering
    \includegraphics[width = \linewidth]{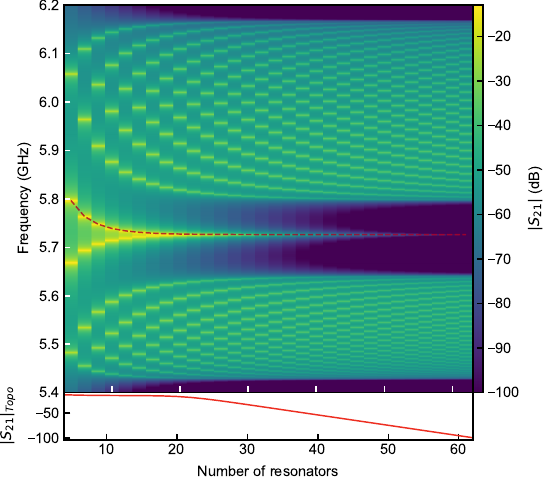}
    \caption{\textbf{Evolution of the CCA transmission in the SSH configuration as a function of the number of cavities.} Input/output simulation using \eq\eqref{app:eq:inputOutput} with Hamiltonian \eq\eqref{methods:hamMat} of the evolution of the amplitude of the CCA transmission, $|S_{21}|$, in a CCA with $J_2/J_1 = 1.57$as a function of the numbers of resonators, $N$. The inset with the red line shows the evolution of the transmission of the symmetric SSH mode as a function of the number of resonators, $N$.}
    \label{fig:topoEvo}
\end{figure}
In \figs\ref{fig:3}\textbf{d} and \textbf{f} one striking observation is represented by the drop in amplitude of the SSH modes for CCAs measured in transmission ($S_{21}$) as a function of the number of resonators, $N$. 
Here, we model this effect and show how it scales using the input/output formalism introduced in \app\ref{app:Dissipations} (\eq\eqref{app:eq:inputOutput}). 
In \fig\ref{fig:topoEvo}, we report the simulated $|S_{21}|$ in an SSH CCA
characterized by $J_2/J_1 = 1.57$ as a function of the number of cavities, $N$.
We clearly observe that, the amplitude of the SSH modes drops drastically as a function of $N$ until the SSH modes are no more visible in transmission.
This effect can be intuitively explained from the nature of the SSH modes.
Indeed, due to their exponential localization at the edges of the CCA, their interaction occurs primarily through the overlap of the tails of their wavefunctions within the bulk region.
As the bulk size increases, the overlap between these modes diminishes, leading to a reduction in hybridization between them. Consequently, the effective photon hopping rate at the frequency of the SSH modes, originating from the coupling of the two microwave ports on the side of the CCA, decreases, resulting in lower transmission. We anticipate a more rapid decrease in transmission with respect to the number of resonators, denoted as $N$, in the strongly localized configuration.

\section{\label{app:nonIdealities} Disorder}

\subsection{Influence of resonator frequency scattering on the bulk modes}
\begin{figure}
    \centering
    \includegraphics[width = \linewidth]{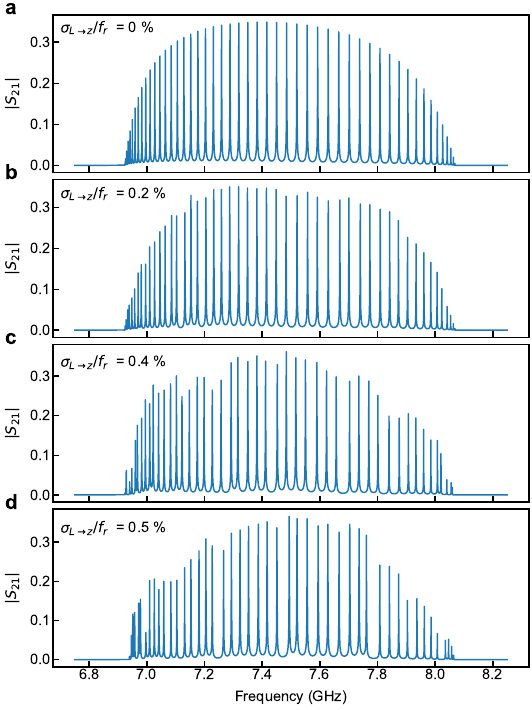}
    \caption{\textbf{Simulation of the transmission, $|S_{21}|$, in a CCA with Gaussian disorder applied to the inductances.} From \textbf{a.} to \textbf{d.}, transmission through CCAs with $N = 50$ resonators ($M=1$) with disorder values of $\sigma_{L\rightarrow z}/f_r = 0\%$ (\textbf{a}),  $\sigma_{L\rightarrow z}/f_r = 0.2 \%$ (\textbf{b}),  $\sigma_{L\rightarrow z}/f_r = 0.4\%$ (\textbf{c}) and  $\sigma_{L\rightarrow z}/f_r = 0.5 \%$ (\textbf{d}), applied to the inductances of the resonators. The simulation is performed using Hamiltonian \eq\eqref{methods:hamMat} using the parameters of the device presented in \fig\ref{fig:3}\textbf{b} and in Tab.~\ref{tab:paramsFig2}.}
    \label{app:fig:Ripples}
\end{figure}
A qualitative way to understand how disorder affects the spectrum of the CCA can be implemented by simulating the CCA transmission, $|S_{21}|$, while introducing scattering on the resonant frequencies induced by Gaussian noise on the inductances, $\sigma_L$. 
In \fig\ref{app:fig:Ripples}, we plot a simulation of the transmission amplitude, $|S_{21}|$, in a uniform CCA with $N = 50$. 
The CCA parameters are the one extracted from the CCA in \fig\ref{fig:2}\textbf{b} for different values of Gaussian noise $\sigma_L$.
In \fig\ref{app:fig:Ripples} we refer to the applied noise as, $\sigma_{L\rightarrow z}$, the disorder induced by $\sigma_L$ on the resonant frequency of the cavities in the CCA. 
When no disorder is applied, the amplitude of the modes follows the same trend as the eigenvalues of the complex part of the non-Hermitian Hamiltonian\eq\eqref{app:eq:nonHermtianHam}. 
As we increase the disorder, we observe the appearance of ripples in the mode's amplitude and frequency deviation from what is reported in \fig\ref{app:fig:Ripples}\textbf{a}. 
This is due to the fact that the resonators are not degenerate anymore. 
From this simulated trend of the transmission spectrum, we can safely establish that the disorder in our devices is below $\sigma_{L\rightarrow z}/f_r = 0.4 \%$.\\

\subsection{Influence of disorder on the SSH modes}
\begin{figure}
    \centering
    \includegraphics[width=\linewidth]{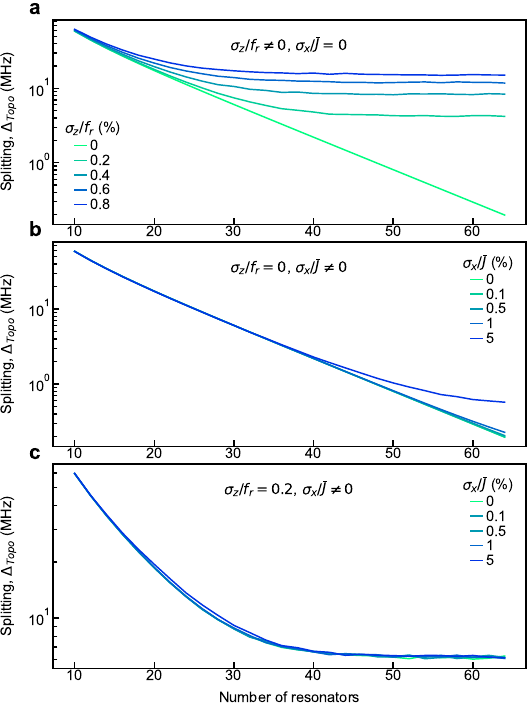}
    \caption{\textbf{Simulation of the influence of $\sigma_z$ and $\sigma_x$ disorder on the SSH modes splittings, $\Delta_\text{Topo}$.} The simulations are performed with a Hamiltonian on a CCA in the SSH configuration with $J_2/J_1 = 1.22$. \textbf{a.} Splittings of the SSH modes, $\Delta_\text{Topo}$, as a function of the number of resonators, $N$, for different values of $\sigma_z$ disorder ($\sigma_x = 0$). \textbf{b.}  $\Delta_\text{Topo}$ as a function of $N$ for different values of $\sigma_x$ ($\sigma_z = 0$). \textbf{c.}  $\Delta_\text{Topo}$ as a function of the $N$ for different values of $\sigma_x$ disorder ($\sigma_z = 0.2\% f_r$).}
    \label{app:fig:SSHDisorder}
\end{figure}

\begin{figure*}
    \centering
    \includegraphics[width=\linewidth]{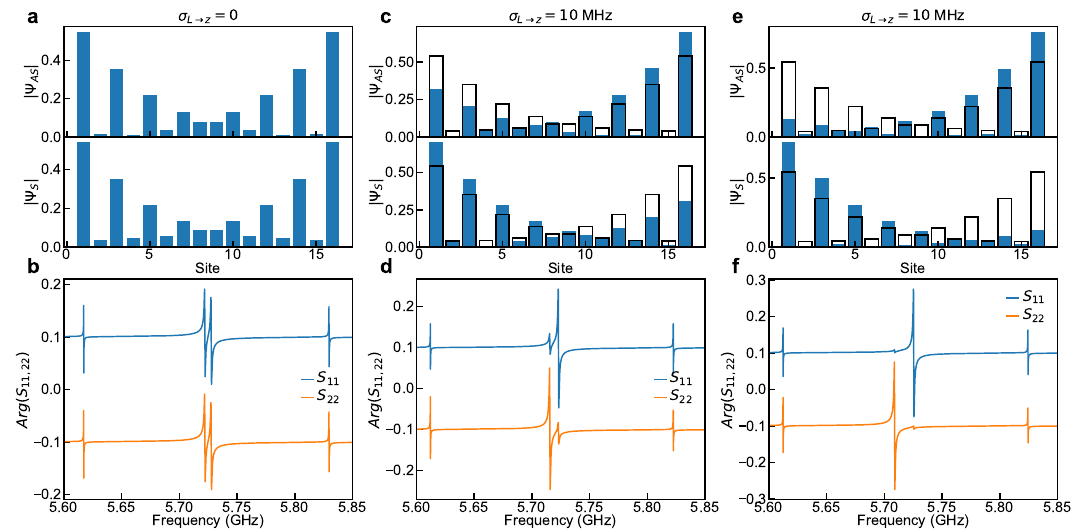}
    \caption{\textbf{Simulation of the influence of disorder on the SSH modes localization.} \textbf{a.} (\textbf{c.}, \textbf{e.}) Simulated mode distribution of the norm of the antisymmetric ($\left|\Psi_{AS}\right|$) and symmetric ($\left|\Psi_{S}\right|$) SSH modes, for different disorder realizations. The black boxes in \textbf{c} and \textbf{e} represent the mode distribution in the disorder-free case \textbf{a}. \textbf{b.} (\textbf{d.},\textbf{f.}) Simulation of Arg($S_{11}$) and Arg($S_{22}$) reflection spectrum for the same disorder realization as in \textbf{a} (\textbf{c}, \textbf{e}). The simulations are performed on CCAs with $N=16$ in the same configuration as the device presented in \fig\ref{fig:3}\textbf{f} of the main text.}
    \label{app:fig:disSSH}
\end{figure*}

Since the SSH modes are protected by chiral symmetry, they are very sensitive to the chiral symmetry breaking terms in Hamiltonian \eq\eqref{eq:k-spaceHam}.
As discussed in the main text, terms proportional to $\tau_z$ will break topological protection.
Such terms simply appear in the Hamiltonian due to frequency scattering between the resonators, which naturally arise due to fabrication imperfection. 
In \fig\ref{app:fig:SSHDisorder}, we simulate the effect of disorder on the SSH modes' hybridization, $\Delta_\text{Topo}$, as a function of the number of resonators, $N$.
The simulation is performed using the Hamiltonian \eq\eqref{methods:hamMat} in Methods, using the parameters of the device in the SSH configuration with $J_2/J_1 = 1.22$.
The $\sigma_L$ noise applied to the inductors induces both $\tau_z$ and $\tau_x$ type of disorder, impacting respectively the resonant frequency and the coupling of the resonators in the CCA.
Although $\tau_z$-type disorder breaks chiral symmetry, we anticipate that the SSH modes will still exhibit some degree of resilience against $\tau_x$-type disorder.

We study three cases: 1) $\sigma_z \neq 0$ and $\sigma_x = 0$ (\fig\ref{app:fig:SSHDisorder}\textbf{a}), 2) $\sigma_z = 0$ and $\sigma_x \neq 0$ (\fig\ref{app:fig:SSHDisorder}\textbf{b}), and 3) $\sigma_z \neq 0$ and $\sigma_x \neq 0$ (\fig\ref{app:fig:SSHDisorder}\textbf{c}).

In \fig\ref{app:fig:SSHDisorder}\textbf{a}, we report the median of $\Delta_\text{Topo}$ for different values of $\sigma_z$. 
For $\sigma_z = 0$, the disorder-free case, one expects $\Delta_\text{Topo}$ to decay exponentially as a function $N$ following \eq\eqref{eq:deltaTopo}. 
By introducing and increasing $\sigma_z$ disorder, we observe a saturation of $\Delta_\text{Topo}$ as a function of the number of cavities.
This saturation value of $\Delta_\text{Topo}$ increases non-uniformly as a function of disorder $\sigma_z$.

In \fig\ref{app:fig:SSHDisorder}\textbf{b}, we implement the same study as above but for $\tau_x$ type disorder.

We can observe a considerable deviation from the expected noiseless case just for high enough values of $\sigma_x$ and $N$.

In \fig\ref{app:fig:SSHDisorder}\textbf{c}, we apply both types of disorder at the same time.
We keep the disorder $\sigma_z$ constant but sweep $\sigma_x$.
As expected, we observe a saturation of $\Delta_\text{Topo}^N$ as a function of $N$, which is independent from the magnitude of $\sigma_x$.
This justifies why in the main text, we focus only on $\tau_z$ type disorder. 
The disorder on the inductance affects not only the SSH modes splitting but also the spatial profile of the SSH modes as can be observed in \fig\ref{app:fig:disSSH}.
Here, the effect of $\sigma_L$ disorder is studied by employing Hamiltonian \eq\eqref{methods:hamMat} for $N = 16$ cavities with $J_2/J_1 = 1.57$.
In a disorder-free case, we expect the SSH modes to hybridize and form a symmetric and an antisymmetric superposition, presenting equal norm on the different CCA site ($|\Psi_S| = |\Psi_{AS}|$, see \fig\ref{app:fig:disSSH}\textbf{a}). 
This symmetry is also observed in the simulation according to \eq\eqref{app:eq:inputOutput} introduced in \app\ref{app:Dissipations} of the phase of the reflected signal at the microwave ports, $\text{Arg}(S_{11},S_{22})$ (see \fig\ref{app:fig:disSSH}\textbf{b}).
From both ports, we find equal phase shifts for the symmetric and antisymmetric SSH modes.

In the remaining panels of \fig\ref{app:fig:disSSH}, we report two independent realization of disordered scenarios of the SSH modes with  $\sigma_{L\rightarrow z} = \SI{10}{\mega\hertz}$.
In \fig\ref{app:fig:disSSH}\textbf{c} and \textbf{e}, we show $|\Psi_{AS}|$ and ($|\Psi_S|$) for the two instances of disorder, respectively.
We observe that the disorder will randomly change the localization of the modes, making them asymmetric.
This behavior is again reflected in $\text{Arg}(S_{11},S_{22})$, as reported in \fig\ref{app:fig:disSSH}\textbf{d} and \fig\ref{app:fig:disSSH}\textbf{f}, respectively corresponding to SSH mode profile in \figs\ref{app:fig:disSSH}\textbf{c} and \textbf{e}.
We can observe a clear asymmetry in the phase shifts from the two microwave ports, which suggest a stronger localization of both SSH modes, on one side or the other of the CCA, as opposed to the disorder-free case (\fig\ref{app:fig:disSSH}\textbf{b}).

\subsection{Influence on the parameter estimation}

\begin{figure*}
    \centering
    \includegraphics[width = \linewidth]{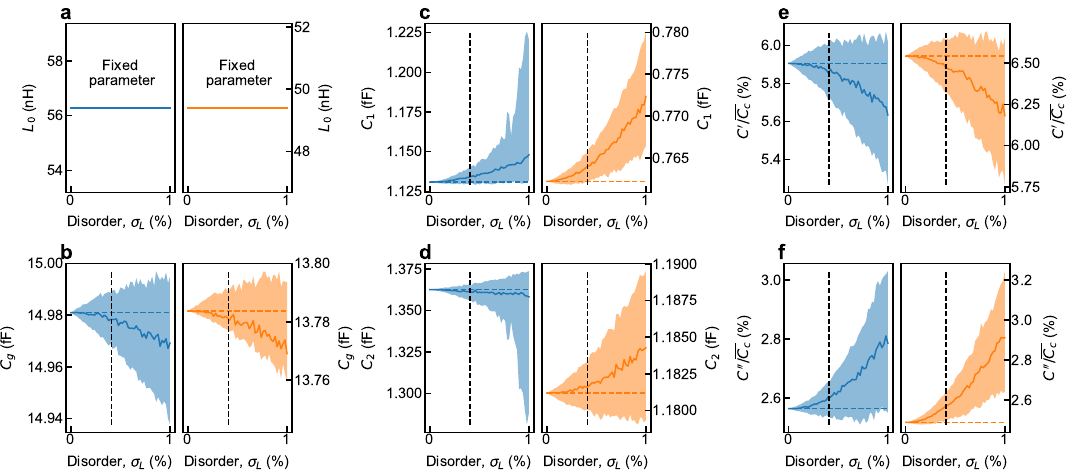}
    \caption{\textbf{Simulation of the influence of disorder on the estimation of parameters.} Statistical simulation of the influence of disorder on the inductance $\sigma_L$, for a $N = 32$ CCA, on the estimation of \textbf{a.}, the inductance to ground, $L_g$, \textbf{b.}, the capacitance to ground $C_g$, \textbf{c.}, the intracell capacitance $C_1$, \textbf{d.}, the intercell capacitance $C_2$, \textbf{e.}, the second nearest neighbor coupling capacitance ratio $C^{\prime}/\overline{C_c}$  and \textbf{f.}, the third nearest neighbor coupling capacitance ratio $C^{\prime\prime}/\overline{C_c}$. The left (blue) and right (orange) panels of each subfigure represent the weak and strong coupling configuration $J_2/J_1 = 1.22$ and $J_2/J_1 = 1.57$, respectively, as in \fig\ref{fig:3}. The continuous lines represent the median of the extracted parameters, and the shaded areas represent the $1\sigma$ uncertainty. The horizontal dashed lines represent the initial parameters set in the simulation.}
    \label{app:fig:paramMis}
\end{figure*}

The estimation of the parameters extracted from the spectra of the CCAs using the model described in Methods \sect\ref{sec:methods} can be affected by disorder.
In order to get an estimation of the error on the extracted parameters, we perform a statistical analysis of the fitting method implemented as explained here below (\fig\ref{app:fig:paramMis}).

We simulate using the Hamiltonian \eq\eqref{methods:hamMat} of a CCA with $N = 32$ cavities for the two SSH coupling configurations introduced in the main text ($J_2/J_1 = 1.22$ and $J_2/J_1 = 1.57$).
We introduce a frequency scattering on the cavities by applying a Gaussian noise, $\sigma_L$, on the inductances of the resonators.
For each value of disorder $\sigma_L$, we perform 500 fits.
In \fig\ref{app:fig:paramMis}, we report the fitted parameters as a function of $\sigma_L$.
For all parameters except the inductance, which is a fixed-fitting parameter, we observe a deviation from the initial parameters of the simulated CCAs in absence of disorder.
For the typical disorder extracted in our study $\sigma_{L}/L_g = 0.44_{-0.06}^{+0.09}$ (highlighted by the black dashed lines in \figs\ref{app:fig:paramMis}\textbf{b}, \textbf{c} and \textbf{d}), we find an error of approximately $\SI{10}{\atto\farad}$ for the capacitances and $0.1 \%$ for the capacitance ratios (\figs\ref{app:fig:paramMis}\textbf{e} and \textbf{f}).

\section{Time-domain measurements \label{app:timedomain}}

In this section, we describe how the time-resolved measurements of the SSH modes are performed and analyzed.

As described in the main text, the measurement is implemented by sending a Gaussian pulse from one of the edges of the CCA at a frequency in the middle of the two SSH modes.
The signal is acquired throughout the full pulse sequence, i.e. before and after the excitation pulse, from both sides of the CCA.
It is then demodulated at the frequency at which the pulse is sent.
The demodulated reflected or transmitted signal, away from the pulse, is expected to take the form~\cite{eichlerClassicalQuantumParametric}
\begin{align}
    |S_{11}|(t) &= e^{-\kappa_1 t}|\cos{\left(gt+\varphi_1\right)}|,\\
    |S_{21}|(t) &= e^{-\kappa_2 t}|\sin{\left(gt+\varphi_2\right)}|,
\end{align}
with $\kappa_1$ and $\kappa_2$ being the total dissipations of SSH modes 1 and 2.
We use this equation to fit the beating profiles presented in \fig\ref{fig:4}\textbf{c} of the main text.

\begin{figure}
    \centering
    \includegraphics[width=\linewidth]{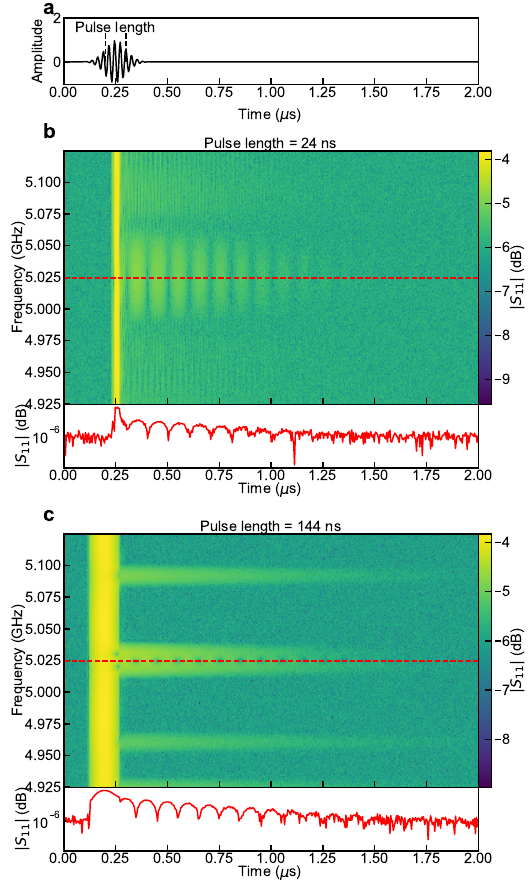}
    \caption{\textbf{Calibration of the time resolved measurements.} \label{app:timeDomain:calPulse} \textbf{a.} Gaussian pulse sent for the calibration measurement. \textbf{b.} (Top) Time-resolved measurement as a function of the frequency of the applied pulse, for a pulse length of \SI{24}{\nano\second} on a CCA with 26 cavities in the configuration $J_2/J_1 = 1.22$. (Bottom) Line-cut at the position of the red dashed line (Top). \textbf{c.} (Top) Time-resolved measurement as a function of the frequency of the applied pulse, for a pulse length of \SI{144}{\nano\second} on a CCA with 26 cavities in the configuration $J_2/J_1 = 1.22$. (Bottom) Line-cut at the position of the red dashed line (Top).}
\end{figure}

The pulse width and shape need to be carefully calibrated to avoid spurious frequency components exciting the other modes of the CCA.
In \fig\ref{app:timeDomain:calPulse}, we show two cases of a Gaussian pulse applied to a CCA with $N=26$ and $J_2/J_1 = 1.22$ (shown in \fig\ref{fig:4} of the main text) for different pulse length of \SI{24}{\nano\second}(\fig\ref{app:timeDomain:calPulse}\textbf{b}) and \SI{144}{\nano\second} (\fig\ref{app:timeDomain:calPulse}\textbf{c}) as a function of the pulse frequency.
The beating pattern developed by the shorter pulse is larger because its frequency components can excite the SSH modes even when detuned from the midpoint of the two SSH modes.
For the pulse with longer length, the beating pattern is only happening, as expected, between the SSH modes.

\section{Extra data}

In this section, we show measurements of some extra devices similar to the one shown in the main text.

\begin{figure}
    \centering
    \includegraphics[width = \linewidth]{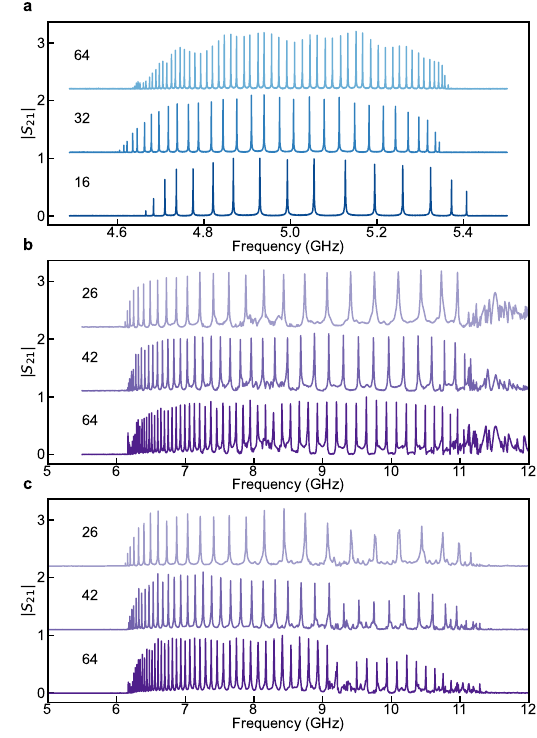}
    \caption{\textbf{Extra spectra of CCAs with $M=1$} \textbf{a.} Transmission measurements, $|S_{21}|$, from a uniform rectangular CCA ($M=1$) with $N = 16,32,64$ and $J/2\pi = \SI{180}{\mega\hertz}$. \textbf{b} Transmission measurements, $|S_{21}|$, from a uniform hexagonal CCA ($M=1$) with $N = 26,42,64$ and $J/2\pi = \SI{1200}{\mega\hertz}$. Partially reproduced in \fig\ref{fig:2}\textbf{c}. \textbf{b.} Same device measured in a different cooldown, with different sample packaging.}
    \label{app:fig:deviceNotMeas}
\end{figure}

In \fig\ref{app:fig:deviceNotMeas}, we show devices associated with \fig\ref{fig:2}\textbf{c}.
In \fig\ref{app:fig:deviceNotMeas}\textbf{a}, we display spectra of devices with $M = 1$ and $N = 16, 32$ and $64$ cavities with the rectangular design.

\figs\ref{app:fig:deviceNotMeas}\textbf{b} and \textbf{c}, exhibit several measurements of CCAs featuring hexagonal geometry with $M = 1$ and $N = 26, 42$ and $64$.
In panel \textbf{b}, we can observe multiple spurious modes starting from \SI{11}{\giga\hertz} on three distinct CCAs.
In panel \textbf{c}, we report the amplitude of the transmission, $|S_{21}|$ of the same device measured during a separate cooldown.
Here, while the additional modes at \SI{11}{\giga\hertz} are no longer visible, other spurious effects emerge around \SI{9.5}{\giga\hertz}.
These observations lead us to conclude that these spurious modes are not intrinsic to the devices themselves.

\begin{figure*}
    \centering
    \includegraphics[width = \linewidth]{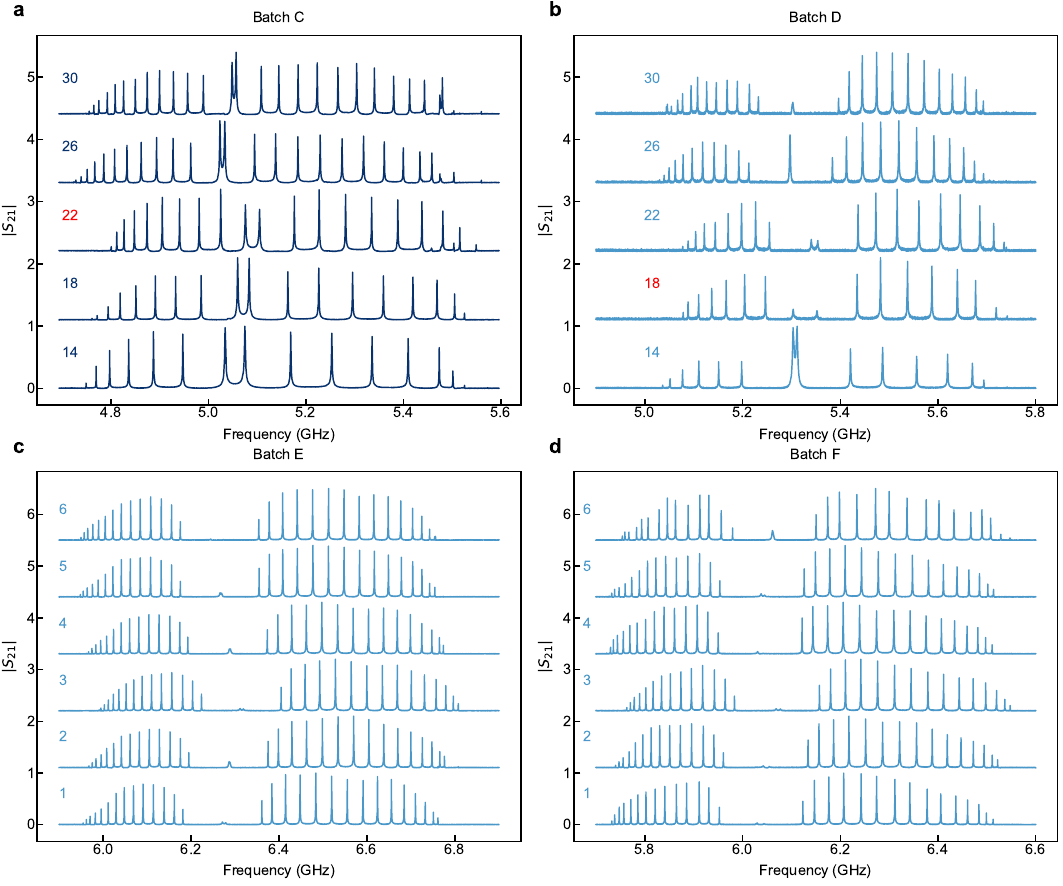}
    \caption{\textbf{Spectrum of SSH CCA used in \fig\ref{fig:4}} \textbf{a.} Amplitude of the CCA transmission, $|S_{21}|$, of devices of batch C. \textbf{b.} Amplitude of the CCA transmission, $|S_{21}|$, of devices of batch D. \textbf{c.} Amplitude of the CCA transmission, $|S_{21}|$, of devices of batch E. \textbf{d.} Amplitude of the CCA transmission, $|S_{21}|$, of devices of batch F. The strong and light blue colors highlight devices in the configuration $J_2/J_1 = 1.22$ and $J_2/J_1 = 1.57$ configurations, respectively. The CCAs in batches C and D are fabricated with a different number of resonators ($N = 14, 18, 22, 26, 30$). Batch E and F consist of 6 repetitions of identical CCA with $N = 32$. The red labels highlight spectra of CCAs presenting strong local disorder identified according to time-resolved measurements (see \sect\ref{sec:disorder} of the main text).}
    \label{app:fig:topoFig4}
\end{figure*}

In \fig\ref{app:fig:topoFig4}, we show $|S_{21}|$ for all the SSH-CCA used in the disorder analysis reported in \fig\ref{fig:4}.

\section{Measurement setup {\label{app:measSetup}}}

\begin{figure*}
    \centering
    \includegraphics[width = \linewidth]{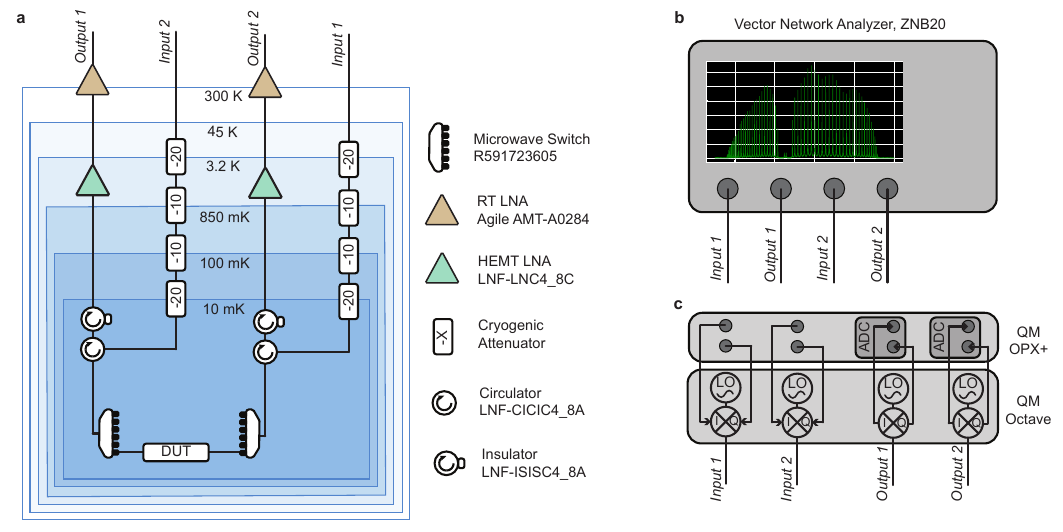}
    \caption{\textbf{Schematic of the measurement setup.} \textbf{a.} Cabling of the cryogenic setup. The input lines are attenuated with 60 dB cryogenic attenuators distributed along the different stages of the cryostat. \textbf{b.} Cabling of the room-temperature Vector Network Analyzer (VNA). The input/output lines correspond to the one of panel \textbf{a}. \textbf{c.} Cabling of the room-temperature time-domain setup.  The input/output lines correspond to the one of panel \textbf{a}.}
    \label{app:fig:measSetup}
\end{figure*}

A schematic of the measurement setup is depicted in \fig\ref{app:fig:measSetup}. 
A simplified version of the cryogenic setup is presented in \fig\ref{app:fig:measSetup}\textbf{a}. 
The device is thermally anchored to the mixing chamber plate of a commercial dry dilution cryostat (Bluefors-LD) at a temperature of $\SI{10}{\milli\kelvin}$. 
Multiple devices can be measured in a single cooldown both in reflection and transmission making use of cryogenic circulators (LNF-CICI48A) and cryogenic switches (Radiall R577432000). 
The input lines are attenuated with cryogenic attenuators at different stages of the cryostat, as reported in \fig\ref{app:fig:measSetup}\textbf{a}. 
The signal output from the device goes through one circulator and three isolators before being amplified at 4K with a HEMT amplifier (LNF-LNC4-8C). 
The signal is then further amplified at room temperature with a low-noise amplifier (Agile AMT-A0284). 
The full scattering matrix of the devices is characterized using an R\&S ZNB20 vector network analyzer (VNA) (Fig.~\ref{app:fig:measSetup}\textbf{b}).\\
We used arbitrary waveform generators (AWG) and Digitizers from an OPX+ from Quantum Machine, for implementing time-resolved measurements. 
This measurement consists of sending a Gaussian pulse at an intermediate frequency from the AWG OPX+ (see Fig.~\ref{app:fig:measSetup}\textbf{c}); it is then up-converted at room temperature with an IQ mixer in the Octave (Quantum-machine) and sent down to the sample. 
After amplification, the reflected and transmitted signals are down-converted by an IQ mixer in the Octave (Quantum-machine) and then digitized in the OPX+ module.
Finally, the signal is demodulated to DC in slices of $\SI{4}{\nano\second}$ and averaged over 20,000 repetitions.

\section{\label{app:paramTable} Tables of parameters}

\begin{table*}
    \centering
    \caption{Table of extracted CCA parameters for \fig\ref{fig:2}. The parameters are defined as follow: $N$ is the number of cavities in the CCA. $\Vec{L_g}$ is the inductance or inductances to ground if $M>2$. $\Vec{C_g}$ is the capacitance to ground or capacitances to ground if $M>2$. $\Vec{C_c}$ is the coupling or coupling capacitances if $M>1$. $\omega_r/2\pi$ is the resonant frequency of the cavities. $Z_r$ is the estimated impedance of the cavities in the array. $\Vec{J}$ is the coupling or couplings between the cavities if $M>1$. $C^\prime/\bar{C_c}$ is the second nearest neighbor capacitance over $\overline{C_c}$. $C^{\prime\prime}/\bar{C_c}$ is the third nearest neighbor capacitance over $\overline{C_c}$.}
    \rowcolors{2}{gray!25}{white}
    \begin{tabular}{c|c|c|c|c|c|c|c|c|c}
        Panel & $N$ & $\Vec{L_g}$ (\SI{}{\nano\henry}) & $\Vec{C_g}$ (\SI{}{\femto\farad}) & $\Vec{C_c}$ (\SI{}{\femto\farad}) & $\omega_r/2\pi$ (\SI{}{\giga\hertz})& $Z_r$ (\SI{}{\kilo\ohm}) & $\Vec{J}$ (\SI{}{\mega\hertz}) & $C^\prime/\bar{C} ~(\%)$ & $C^{\prime\prime}/\bar{C} ~(\%)$\\ 
        \hline\hline
         & 25 & 30.25 & 13.02 & 1.14 & 7.43 & 1.52 & 280 & 5.98 & 1.24 \\
        \textbf{b} & 50 & 30.25 & 13.02 & 1.14 & 7.43 & 1.52 & 280 & 5.91 & 1.16 \\
         & 100 & 30.25 & 13 & 1.14 & 7.43 & 1.52 & 280 & 5.88 & 1.23 \\
         
         \hline
         
         & 51 & 65.8 & 16.09 & 0.4 & 4.77 & 2.02 & 57 & 6.73 & 2.03 \\
        \textbf{c} & 64 & 65.8 & 12 & 0.8 & 5.32 & 2.34 & 164 & 4.9 & 1.89\\
         & 64 & 18.8 & 10.9 & 6.09 & 8.09 & 0.9 & 1164 & 0.01 & 0.01\\
         
         \hline
         
        \multirow{2}{*}{\textbf{d}} & 32 & 56.56 & 16.36 & \Gape[0pt][2pt]{\makecell{1.48\\1.24}} & 4.86 & 1.72 & \Gape[0pt][2pt]{\makecell{191\\160}} & 5.91 & 2.03 \\
         & 32 & 48.104 & 14.02 & \Gape[0pt][2pt]{\makecell{1.20\\0.77}} & 5.75 & 1.73 &  \Gape[0pt][2pt]{\makecell{217\\141}} & 6.24 & 2.57\\
         
        \hline
        
         & 60 & \Gape[0pt][2pt]{\makecell{46.43\\44.05}} & \Gape[0pt][2pt]{\makecell{13.93\\14.03}} & \Gape[0pt][2pt]{\makecell{1.19\\1.19\\0.76}} & 5.874 & 1.7 & \Gape[0pt][2pt]{\makecell{214\\226\\144}} & 5.99 & 1.33 \\
        \textbf{e} & 56 & \Gape[0pt][2pt]{\makecell{46.43\\48.88}} & \Gape[0pt][2pt]{\makecell{13.63\\13.78}} & \Gape[0pt][2pt]{\makecell{1.03\\1.03\\0.86\\1.03}} & 5.91 & 1.77 & \Gape[0pt][2pt]{\makecell{197\\168\\164\\160}} & 5.31 & 0.76 \\
         & 60 & \Gape[0pt][2pt]{\makecell{46.43\\44.05}} & \Gape[0pt][2pt]{\makecell{13.36\\14.83}} & \Gape[0pt][2pt]{\makecell{1.16\\1.16\\1.16\\1.16\\0.81}} & 5.91 & 1.77 & \Gape[0pt][2pt]{\makecell{200\\192\\192\\190\\164}} & 6.02 & 0.68\\
    \end{tabular}
    \label{tab:paramsFig2}
\end{table*}

\begin{table*}
    \centering
    \caption{Table of extracted CCA parameters for \fig\ref{fig:3}. The parameters are defined as follow: $N$ is the number of cavities in the CCA. $\Vec{L_g}$ is the inductance or inductances to ground if $M>2$. $\Vec{C_g}$ is the capacitance to ground or capacitances to ground if $M>2$. $\Vec{C_c}$ is the coupling or coupling capacitances if $M>1$. $\omega_r/2\pi$ is the resonant frequency of the cavities. $Z_r$ is the estimated impedance of the cavities in the array. $\Vec{J}$ is the coupling or couplings between the cavities if $M>1$. $C^\prime/\bar{C_c}$ is the second nearest neighbor capacitance over $\overline{C_c}$. $C^{\prime\prime}/\bar{C_c}$ is the third nearest neighbor capacitance over $\overline{C_c}$.}
    \rowcolors{2}{gray!25}{white}
    \begin{tabular}{c|c|c|c|c|c|c|c|c|c}
        Panel/Batch & $N$ & $\Vec{L_g}$ (\SI{}{\nano\henry}) & $\Vec{C_g}$ (\SI{}{\femto\farad}) & $\Vec{C_c}$ (\SI{}{\femto\farad}) & $\omega_r/2\pi$ (\SI{}{\giga\hertz})& $Z_r$ (\SI{}{\kilo\ohm}) & $\Vec{J}$ (\SI{}{\mega\hertz}) & $C^\prime/\bar{C} ~(\%)$ & $C^{\prime\prime}/\bar{C} ~(\%)$ \\ 
        \hline\hline
         & 16 & 56.28 & 14.94 & \makecell{1.13\\1.35} & 5.1 & 1.8 & \makecell{168\\200} & 6.04 & 2.32\\
        \textbf{d}/A & 32 & 56.28 & 14.99 & \Gape[0pt][2pt]{\makecell{1.13\\1.37}} & 5.09 & 1.79 & \Gape[0pt][2pt]{\makecell{166\\201}} & 5.78 & 2.88\\

         & 64 & 56.28 & 15 & \Gape[0pt][2pt]{\makecell{1.13\\1.36}} & 5.09 & 1.79 & \Gape[0pt][2pt]{\makecell{166\\200}} & 5.89 & 2.47\\
         \hline
         & 16 & 49.39 & 13.64 & \Gape[0pt][2pt]{\makecell{0.75\\1.17}} & 5.75 & 1.78 & \Gape[0pt][2pt]{\makecell{141\\218}} & 6.83 & 2.42\\

        \textbf{e}/B & 32 & 49.39 & 13.89 & \Gape[0pt][2pt]{\makecell{0.76\\1.19}} & 5.70 & 1.76 & \Gape[0pt][2pt]{\makecell{140\\215}} & 6.34 & 2.55\\

         & 64 & 49.39 & 13.81 & \Gape[0pt][2pt]{\makecell{0.76\\1.18}} & 5.72 & 1.77 & \Gape[0pt][2pt]{\makecell{141\\216}} & 6.44 & 2.47\\
    \end{tabular}
    \label{tab:paramsFig3}
\end{table*}

\begin{table*}
    \centering
    \caption{Table of extracted CCA parameters for \fig\ref{fig:4}. The parameters are defined as follow: $N$ is the number of cavities in the CCA. $\Vec{L_g}$ is the inductance or inductances to ground if $M>2$. $\Vec{C_g}$ is the capacitance to ground or capacitances to ground if $M>2$. $\Vec{C_c}$ is the coupling or coupling capacitances if $M>1$. $\omega_r/2\pi$ is the resonant frequency of the cavities. $Z_r$ is the estimated impedance of the cavities in the array. $\Vec{J}$ is the coupling or couplings between the cavities if $M>1$. $C^\prime/\bar{C_c}$ is the second nearest neighbor capacitance over $\overline{C_c}$. $C^{\prime\prime}/\bar{C_c}$ is the third nearest neighbor capacitance over $\overline{C_c}$.}
    \rowcolors{2}{gray!25}{white}
    \begin{tabular}{c|c|c|c|c|c|c|c|c|c}
        Batch & $N$ & $\Vec{L_g}$ (\SI{}{\nano\henry}) & $\Vec{C_g}$ (\SI{}{\femto\farad}) & $\Vec{C_c}$ (\SI{}{\femto\farad}) & $\omega_r/2\pi$ (\SI{}{\giga\hertz})& $Z_r$ (\SI{}{\kilo\ohm}) & $\Vec{J}$ (\SI{}{\mega\hertz}) & $C^\prime/\bar{C} ~(\%)$ & $C^{\prime\prime}/\bar{C} ~(\%)$ \\ 
        \hline\hline
         & 14 & 56.22 & 14.92 & \Gape[0pt][2pt]{\makecell{1.18\\1.41}} & 5.09 & 1.79 & \Gape[0pt][2pt]{\makecell{174\\207}} & 5.17 & 0.77\\

         & 18 & 56.22 & 14.82 & \Gape[0pt][2pt]{\makecell{1.16\\1.41}} & 5.11 & 1.79 & \Gape[0pt][2pt]{\makecell{173\\209}} & 4.72 & 0.9\\

        C & 22 & 56.22 & 14.67 & \Gape[0pt][2pt]{\makecell{1.21\\1.33}} & 5.13 & 1.81 & \Gape[0pt][2pt]{\makecell{182\\200}} & 5.78 & 0.5\\

         & 26 & 56.22 & 15.06 & \Gape[0pt][2pt]{\makecell{1.18\\1.44}} & 5.06 & 1.78 & \Gape[0pt][2pt]{\makecell{171\\208}} & 5.51 & 1.24\\

         & 30 & 56.22 & 14.95 & \Gape[0pt][2pt]{\makecell{1.17\\1.42}} & 5.09 & 1.79 & \Gape[0pt][2pt]{\makecell{172\\208}} & 5.29 & 0.74\\
         \hline
         & 14 & 56.05 & 13.98 & \Gape[0pt][2pt]{\makecell{0.78\\1.20}} & 5.33 & 1.87 & \Gape[0pt][2pt]{\makecell{131\\202}} & 5.69 & 0.44\\

         & 18 & 56.05 & 13.81 & \Gape[0pt][2pt]{\makecell{0.9\\1.07}} & 5.36 & 1.88 & \Gape[0pt][2pt]{\makecell{155\\184}} & 6.29 & 0\\

        D & 22 & 56.05 & 13.77 & \Gape[0pt][2pt]{\makecell{0.76\\1.18}} & 5.37 & 1.88 & \Gape[0pt][2pt]{\makecell{133\\204}} & 5.6 & 0.4\\

         & 26 & 56.05 & 14.01 & \Gape[0pt][2pt]{\makecell{0.78\\1.21}} & 5.33 & 1.87 & \Gape[0pt][2pt]{\makecell{132\\203}} & 6.13 & 1.16\\

         & 30 & 56.05 & 13.95 & \Gape[0pt][2pt]{\makecell{0.77\\1.20}} & 5.34 & 1.87 & \Gape[0pt][2pt]{\makecell{132\\203}} & 5.75 & 0.72\\
         \hline
          & 32 & 48.10 & 13.84 & \Gape[0pt][2pt]{\makecell{0.76\\1.18}} & 6.07 & 1.66 & \Gape[0pt][2pt]{\makecell{150\\230}} & 6.33 & 2.43\\

         & 32 & 48.10 & 13.78 & \Gape[0pt][2pt]{\makecell{0.76\\1.18}} & 6.08 & 1.67 & \Gape[0pt][2pt]{\makecell{150\\230}} & 6.35 & 2.41\\

        E & 32 & 48.10 & 13.65 & \Gape[0pt][2pt]{\makecell{0.75\\1.17}} & 6.11 & 1.67 & \Gape[0pt][2pt]{\makecell{150\\231}} & 6.41 & 2.37\\

         & 32 & 48.10 & 13.78 & \Gape[0pt][2pt]{\makecell{0.76\\1.18}} & 6.07 & 1.66 & \Gape[0pt][2pt]{\makecell{150\\230}} & 6.36 & 2.41\\

         & 32 & 48.10 & 13.87 & \Gape[0pt][2pt]{\makecell{0.76\\1.19}} & 6.08 & 1.66 & \Gape[0pt][2pt]{\makecell{150\\230}} & 6.25 & 2.43\\

         & 32 & 48.10 & 13.87 & \Gape[0pt][2pt]{\makecell{0.76\\1.18}} & 6.1 & 1.67 & \Gape[0pt][2pt]{\makecell{151\\232}} & 6.58 & 2.28\\
        \hline
         & 32 & 48.10 & 13.84 & \Gape[0pt][2pt]{\makecell{0.76\\1.18}} & 6.31 & 1.60 & \Gape[0pt][2pt]{\makecell{155\\238}} & 6.38 & 2.52\\

         & 32 & 48.10 & 13.79 & \Gape[0pt][2pt]{\makecell{0.76\\1.18}} & 6.32 & 1.61 & \Gape[0pt][2pt]{\makecell{155\\239}} & 6.33 & 2.51\\

        F & 32 & 48.10 & 13.68 & \Gape[0pt][2pt]{\makecell{0.76\\1.17}} & 6.35 & 1.61 & \Gape[0pt][2pt]{\makecell{156\\240}} & 6.33 & 2.54\\

         & 32 & 48.10 & 13.84 & \Gape[0pt][2pt]{\makecell{0.77\\1.19}} & 6.32 & 1.61 & \Gape[0pt][2pt]{\makecell{155\\239}} & 6.32 & 2.52\\

         & 32 & 48.10 & 13.83 & \Gape[0pt][2pt]{\makecell{0.77\\1.19}} & 6.30 & 1.60 & \Gape[0pt][2pt]{\makecell{155\\238}} & 6.19 & 2.56\\

         & 32 & 48.10 & 13.72 & \Gape[0pt][2pt]{\makecell{0.76\\1.18}} & 6.30 & 1.60 & \Gape[0pt][2pt]{\makecell{155\\237}} & 6.26 & 2.72\\
    \end{tabular}
    \label{tab:paramsFig4}
\end{table*}

\end{document}